\documentclass[
  aps,
  prl,
  twocolumn,
  superscriptaddress,
  nofootinbib,
  floatfix,
  amsmath,
  amssymb,
  amsfonts
]{revtex4-2}

\usepackage[utf8]{inputenc}
\usepackage{bm}
\usepackage{graphicx}
\usepackage{array}
\usepackage{makecell}
\usepackage{multirow}
\usepackage{diagbox}
\usepackage{xspace}
\usepackage[dvipsnames]{xcolor}
\usepackage{xparse}
\usepackage{etoolbox}
\usepackage{float}
\usepackage[
  bookmarksnumbered,
  bookmarksopen,
  breaklinks=true,
  colorlinks=true,
  citecolor=blue,
  linkcolor=blue,
  urlcolor=blue
]{hyperref}

\DeclareMathOperator*{\argmax}{argmax}

\def\be{\begin{equation}}
\def\ee{\end{equation}}
\def\bea{\begin{eqnarray}}
\def\eea{\end{eqnarray}}
\newcommand{\bes}{\begin{subequations}}
\newcommand{\ees}{\end{subequations}}
\newcommand{\besa}{\begin{subequations}\begin{align}}
\newcommand{\eesa}{\end{align}\end{subequations}}
\def\comment#1{}

\newcommand{\PRLheading}[1]{%
  \par\phantomsection\noindent{\bfseries\boldmath #1.---}\hspace{0.25em}\ignorespaces}
\newcommand{\fitresult}[2]{\makecell{$#1$\\[-1pt]$#2$}}
\newcommand{\fitresulttwo}[1]{\makecell{$#1$\\[-1pt]($\sim$)}}

\allowdisplaybreaks
\AtBeginEnvironment{table}{\footnotesize\renewcommand{\arraystretch}{0.92}}
\AtBeginEnvironment{table*}{\footnotesize\renewcommand{\arraystretch}{0.92}}

\begin{document}
\raggedbottom
\setcounter{secnumdepth}{3}
\setlength{\abovedisplayskip}{4pt plus 1pt minus 1pt}
\setlength{\belowdisplayskip}{4pt plus 1pt minus 1pt}
\setlength{\abovedisplayshortskip}{2pt plus 1pt}
\setlength{\belowdisplayshortskip}{2pt plus 1pt}
\setlength{\textfloatsep}{7pt plus 2pt minus 2pt}
\setlength{\floatsep}{6pt plus 2pt minus 2pt}

\title{Coherent End-to-End Search for Generic Extreme-Mass-Ratio Inspirals }

\author{Xiaobo Zou}
\affiliation{School of Fundamental Physics and Mathematical Sciences, Hangzhou Institute for Advanced Study, UCAS, Hangzhou 310024, China}
\author{Xingyu Zhong}
\affiliation{School of Fundamental Physics and Mathematical Sciences, Hangzhou Institute for Advanced Study, UCAS, Hangzhou 310024, China}
\author{Wen-Biao Han}
\email{wbhan@shao.ac.cn}
\affiliation{Shanghai Astronomical Observatory, Chinese Academy of Sciences, Shanghai 200030, China}
\affiliation{School of Fundamental Physics and Mathematical Sciences, Hangzhou Institute for Advanced Study, UCAS, Hangzhou 310024, China}
%\affiliation{Taiji Laboratory for Gravitational Wave Universe (Beijing/Hangzhou), University of Chinese Academy of Sciences, Beijing 100049, China}
\affiliation{School of Astronomy and Space Science, University of Chinese Academy of Sciences, Beijing 100049, China}
\author{Soumya D. Mohanty}
\email{Soumya.Mohanty@utrgv.edu}
\affiliation{Department of Physics and Astronomy, The University of Texas Rio Grande Valley, Brownsville, Texas 78520, USA}

\begin{abstract}
Extreme-mass-ratio inspirals (EMRIs) encode more than $10^5$ strong-field orbital cycles and are key targets for space-borne gravitational-wave interferometers, yet coherent recovery of generic systems over astrophysically broad priors remains unresolved. Successive Mock LISA, LISA, and Taiji Data Challenges (MLDCs, LDCs, and TDCs) have not yet produced a complete, generally reliable solution for blind EMRI detection and parameter recovery across such priors. The central obstacle is a needle-in-a-haystack likelihood: six phase-evolution parameters span a vast domain, producing an exceptionally narrow global maximum amid numerous secondary maxima. We show that higher-likelihood secondary maxima concentrate progressively around the global maximum and can therefore guide an adaptive contraction of the search volume. We exploit this structure through a reduced-dimensional profile likelihood and a coherent hierarchical strategy to search for EMRI signals across the full 14-dimensional parameter space. This enables the first end-to-end coherent parameter estimation for generic EMRIs with astrophysically broad priors. In stationary Gaussian LISA noise, the search recovers two half-year analytical-kludge signals with signal-to-noise ratios near 50, yielding fitting factors of 0.989 and 0.971, fractional errors of $10^{-3}$--$10^{-2}$ in the phase-evolution parameters and near $3\%$ in the distance, and error of less than $0.1$ radian in the sky location. The method turns secondary maxima into guides for a coherent hierarchical search.
\end{abstract}
\maketitle

\PRLheading{Introduction}
An extreme-mass-ratio inspiral (EMRI) is a stellar-mass compact object orbiting a massive black hole on a generic eccentric and inclined trajectory~\cite{Amaro-Seoane:2012lgq}. During its final year before plunge, it can complete more than $10^5$ phase-coherent cycles in the millihertz band~\cite{Gair:2004iv}. These long-lived signals act as remarkably precise probes of the central spacetime: they can measure the massive black hole's mass and spin~\cite{Barack:2003fp,Cui:2025bgu}, test general relativity in the strong-field regime~\cite{Berry:2019wgg,Babak:2017tow,Fan:2020zhy,Chua:2018yng,Yang:2019xro,Xin:2018urr,Shen:2023pje,Shen:2025svs,Zi:2021pdp,Barack:2006pq,Speri:2024qak,Zou:2025fsg,Long:2026dcb}, and constrain environmental effects near galactic nuclei~\cite{Destounis:2022obl,Feng:2025fkc}. For this reason, EMRIs are primary targets for LISA~\cite{LISA:2017pwj}, Taiji~\cite{Hu:2017mde}, and TianQin~\cite{TianQin:2015yph}. More broadly, a Taiji--TianQin network can probe supermassive-black-hole formation through massive-black-hole-binary populations~\cite{Shen:2026smbh}. Yet the same long, multiharmonic phase evolution that makes them so informative also makes them unusually difficult to recover. A minute shift in a source parameter can accumulate a large waveform error; detector motion adds sky-dependent delays; and the superposition of harmonics carves a high-dimensional, strongly multimodal likelihood. Over nearly two decades, the Mock LISA, LISA, and Taiji Data Challenges (MLDCs~\cite{Babak:2008aa,MockLISADataChallengeTaskForce:2009wir}, LDCs~\cite{Baghi:2022ucj}, and TDCs~\cite{Du:2025xdq}) have driven steady progress in EMRI analysis. They have fostered time-frequency analysis~\cite{Gair:2008ec,Speri:2025ucn}, phenomenological waveforms~\cite{Wang:2012xh}, harmonic-constrained sampling~\cite{Babak:2009ua}, Markov-chain Monte Carlo with genetic~\cite{Cornish:2008zd} or tempering schemes~\cite{Ali:2012zz}, nonlocal degeneracy~\cite{Chua:2021aah} and one-stop function~\cite{Chua:2022ssg}, semicoherent statistics~\cite{Ye:2023lok,Speri:2025ucn,Wang:2026lcc}, iterative prior reduction~\cite{Ye:2023lok,Strub:2025dfs,Cole:2025sqo,Wang:2026lcc}, global optimization~\cite{Wang:2026lcc,Zou:2024jqv,Zou:2024osb}, and artificial intelligence methods~\cite{Cole:2025sqo,Zhang:2022xuq,Zhao:2023ncy,Yun:2023vwa,Liang:2025vuf,Liang:2026cki}. Recent end-to-end studies have reached identification, prior-volume reduction, or parameter inference from broad priors, but the demonstrated configurations use Schwarzschild models, semicoherent or phenomenological statistics, or source-targeted priors. To our knowledge, the method presented here provides the first coherent search for a generic, spinning EMRI across the complete $14$-dimensional parameter space, with population-scale mass, spin, inclination, eccentricity, and all-sky priors (Sec.~\ref{app:prior-work} of the Supplemental Material). We retain the sensitivity of matched filtering while easing its search burden with seven- and eight-dimensional likelihoods, nested optimization of the remaining source parameters, and particle-swarm optimization (PSO)~\cite{Kennedy:1995,Kennedy:2007}.

Against this backdrop, the central obstacle is not merely evaluating the likelihood but finding its sharp primary maximum across a broad prior. The six parameters $\{\mu,M,\lambda,S/M^2,e_0,\nu_0\}$ that govern the long-term phase evolution span the ranges predicted by EMRI population models~\cite{Babak:2017tow,Fan:2020zhy}; within that large search region, they create a narrow ``needle'' in the likelihood landscape~\cite{Chua:2021aah}. This difficulty naturally calls for a hierarchical search that narrows the region step by step. More specifically, partial matches to different subsets of the waveform harmonics generate numerous secondary maxima. Since the harmonics carry unequal signal-to-noise ratios (SNRs), a template far from the primary maximum matches fewer of them, whereas the higher-log-likelihood-ratio (LLR) secondary maxima gather ever more densely as they approach the primary maximum. We therefore turn this pattern into a search guide: independent PSO runs collect secondary maxima in the seven-dimensional likelihood, peaks above a progressively raised LLR threshold set the contracted prior for the next stage, and a final eight-dimensional search refines the recovered parameters. We test the strategy on two $0.5$-year analytical-kludge signals with target SNR $50$, sampled at $15$-s intervals in stationary Gaussian LISA noise. Both signals are detected in the first stage and localized after eight and six stages, respectively; the final searches yield fitting factors of $0.989$ and $0.971$, fractional errors of $10^{-3}$--$10^{-2}$ in the phase-evolution parameters and near $3\%$ in the distance, and error of less than $0.1$ radian in sky location.

\PRLheading{Methodology}\label{method}
The EMRI signal is described by $14$ source parameters. The search separates parameters controlling phase evolution and detector delays from time-independent parameters describing orientation, initial phase, and amplitude. This separation yields eight- and seven-dimensional profile likelihoods implemented through nested optimization. A hierarchical search strategy, combined with these profile likelihoods, carries out the search. Detector response, noise, and waveform definitions are provided in Sec.~\ref{model} of the Supplemental Material.

%#=============================
\begingroup
\small
\PRLheading{Reduced dimensionality likelihood: $8$D}\label{Reduced-Dim-LLR}
For observed data $\overline{d}^I$ and a template waveform $\overline{h}^I(\Theta)$ in time-delay interferometry (TDI)~\cite{Tinto:2004wu} channel $I\in\{A,E\}$, with $\overline{d}^I=\overline{h}^I+\overline{n}^I$, the log-likelihood ratio (LLR) in stationary Gaussian noise is
%14D
\bea
\Lambda(\Theta) = \sum_{I\in \{A, E\}} \left[-(\overline{h}^I(\Theta)|\overline{h}^I(\Theta)) + 2(\overline{d}^I|\overline{h}^I(\Theta))\right]\;, \nonumber \\ 
\label{eq:14D-LLR}
\eea 
where the noise power spectral density follows the LISA SciRDv1 model~\cite{LDC-code-maunal} (see Sec.~\ref{psd} of the Supplemental Material for details).

%13D
The source distance $D$ only rescales the waveform and does not affect the phase match, so it can be maximized analytically~\cite{Babak:2009ua}. Writing $\overline{h}^I(\Theta)=\mathrm{A}\overline{s}^I(\theta^\prime)$, where $\mathrm{A}=1/D$ and $\theta^\prime=\Theta\setminus\{D\}$, gives
\bea
\widehat{\mathrm{A}} =& \argmax\limits_{\mathrm{A}}\Lambda(\Theta) = \frac{\big[\sum_{I\in \{A, E\}} (\overline{d}^I|\overline{s}^I(\theta^\prime))\big]}{\big[\sum_{I\in \{A, E\}} (\overline{s}^I(\theta^\prime)|\overline{s}^I(\theta^\prime))\big]}\;.
\eea
Substitution of $\widehat{\mathrm{A}}$ gives the profile LLR
\bea
\label{eq:8D-LLR}
\rho^2(\theta^\prime) & = \Lambda(\widehat{\mathrm{A}},\theta^\prime)= \frac{\big[\sum_{I\in \{A, E\}} (\overline{d}^I|\overline{s}^I(\theta^\prime))\big]^2}{\big[\sum_{I\in \{A, E\}} (\overline{s}^I(\theta^\prime)|\overline{s}^I(\theta^\prime))\big]}\;, \\ \nonumber
\eea
where $\rho$ is the square root of the LLR. The global optimum $\theta_{*}$ satisfies
\bea
L_G(\theta_{*}) &= \max\limits_{\theta^\prime} \rho^2(\theta^\prime)\;.
\label{eq:glrt1}
\eea 

The eight parameters
	\bea
    \label{eq:8D_p}
    \theta_{\rm 8D} =  \theta_1 \cup  \theta_2\;, 
    \eea
control the orbital phase evolution and detector delays (defined in Sec.\ref{waveform} of the Supplemental Material). The remaining subset $\theta_3$ is independent of the TDI phase evolution  and can be separated from $\theta_{\rm 8D}$ by decomposing the 13D strain $\overline{s}^{i}_{l}(\theta^\prime)$ for harmonic $i$ and arm $l$:
    \bea
    \label{eq:8D_decomposition_hphc}
	\overline{s}^{i}_{l}(\theta^\prime)=&F_l^{+}(\theta_2,\psi){\rm Re}(\overline{s}^{i}_{+}(\theta^\prime))+F_l^{\times}(\theta_2,\psi){\rm Re}(\overline{s}^{i}_{\times}(\theta^\prime))\;, \nonumber \\   
	=&{\rm Re}(e^{i\Phi^i_0}A^c_{+}(\theta_k,\phi_k,\theta_2,\lambda))F_l^{+}(\theta_2,\psi){\rm Re}(\overline{x}^i(\theta_1))  \nonumber  \\  
	-&{\rm Im}(e^{i\Phi^i_0}A^c_{+}(\theta_k,\phi_k,\theta_2,\lambda))F_l^{+}(\theta_2,\psi){\rm Im}(\overline{x}^i(\theta_1))  \nonumber  \\  
	+&{\rm Re}(e^{i\Phi^i_0}A^c_{\times}(\theta_k,\phi_k,\theta_2,\lambda))F_l^{\times}(\theta_2,\psi){\rm Re}(\overline{x}^i(\theta_1)) \nonumber  \\  
	-&{\rm Im}(e^{i\Phi^i_0}A^c_{\times}(\theta_k,\phi_k,\theta_2,\lambda))F_l^{\times}(\theta_2,\psi){\rm Im}(\overline{x}^i(\theta_1)), \nonumber \\   
	=&\sum_{p=1}^{4}a^{i}_{p} \overline{x}^{i}_{l,p}(\theta_{\rm 8D})\;. \\ \nonumber
\eea
Because TDI delays act only on the time-dependent terms in Eq.~(\ref{eq:8D_decomposition_hphc}), the decomposition remains valid for the $13$D TDI channel $I$ of harmonic $i$:
\bea
\overline{s}^{I,i}(\theta^\prime)&=&\sum_{p=1}^{4}a^{i}_{p} \overline{x}^{I,i}_{p}(\theta_{\rm 8D})\;. 
\label{eq:8D_decomposition_TDI}\\ \nonumber
\eea

Substituting Eq.~(\ref{eq:8D_decomposition_TDI}) into Eq.~(\ref{eq:8D-LLR}) gives
\bea
(\overline{d}^I(\theta^\prime)|\overline{s}^I(\theta^\prime))=\sum_{i=1}^{N}\sum_{p=1}^4 a^i_p (\overline{d}^I\big|\overline{x}^{I,i}_{p}(\theta_{\rm 8D}))\;, \nonumber \\  
(\overline{s}^I(\theta^\prime)|\overline{s}^I(\theta^\prime))=\sum_{i=1}^{N}\sum_{j=1}^{N}\sum_{p=1}^4\sum_{q=1}^4 a^i_p a^j_q (\overline{x}^{I,i}_{p}(\theta_{\rm 8D}) \big| \overline{x}^{I,j}_{q}(\theta_{\rm 8D}))\;. \nonumber \\
\label{eq:13D_innerproduct_full}
\eea
This decomposition separates $\theta_3$ from $\theta_{\rm 8D}$ within the likelihood~\cite{Zou:2024osb}.

\PRLheading{$7$D}
For a discrete signal $\overline{x}(t)$ sampled at intervals $\Delta t$, a shift of $n$ samples in time corresponds to a uniform phase rotation in frequency:
\bea
\overline{x}(t-n\Delta t) = {\rm IFFT}\left[\tilde{x}(f)e^{-i2\pi fn\Delta t}\right]\;.
\eea
The shifted time vector $\overline{t}_{\rm sh}$ starts at $t_0+n\Delta t$, with $t_{\rm sh}=t-n\Delta t$.

For an EMRI signal, $\nu_0$ determines the orbital frequency at the template start time and is therefore degenerate with $t_0$. We fix a fiducial value $\nu_0^{\rm fiducial}$ at $t_0$ and define the seven-dimensional parameter set
\bea
\theta_{\rm 7D} =  \theta_2  \cup \theta_1  \setminus \{\nu_0\} \;.
\eea
For each $\theta_{\rm 7D}$, we compute $\nu(t)$ and the $8$D TDI template $\overline{s}^I(t)$ from $t_0$. We then shift $\overline{s}^I(t)$ by $n_{*}\Delta t$ to maximize Eq.~(\ref{eq:8D-LLR}):
\bea
n_{*} =& \argmax\limits_{n}\rho^2(n|\{\theta_{\rm 7D}, \theta_3\})\;.
\eea
This is equivalent to
\bea
\label{eq:7D-1}
n_{*} =& \argmax\limits_{n}\left|\sum_{I\in \{A, E\}}(\overline{d}^I(t)|\overline{h}^I(t_{\rm sh}))\right|\;, \nonumber \\  
=& \argmax\limits_{n}2\left| \sum_{t} \sum_{I\in \{A, E\}}\overline{d}^I_{\rm dw}(t)*\overline{h}^I(t_{\rm sh})\right|\Delta t \;, \nonumber   \\  
\eea 
because the self-inner product in Eq.~(\ref{eq:13D_innerproduct_full}) is invariant under the lag shift. Here Parseval's theorem is used, and $\overline{d}^I_{\rm dw}(t)=\mathrm{IFFT}(\widetilde{d^I_{\rm dw}}/S_n(f))$. The convolution theorem then evaluates Eq.~(\ref{eq:7D-1}) efficiently:
\bea
\label{eq:lag-shift}
n_{*} =& 
\argmax\limits_{n}\left|\mathrm{\sum_{I\in \{A, E\}}\mathrm{cov} (\overline{d}^I_{\rm dw}(t), \overline{h}^I(t))}\right|\;.   \nonumber \\ 
\eea 
The corresponding estimate of $\nu_0$ is $\nu(t_0+n_{*}\Delta t)$.

%The choice of fiducial $\nu_0$ in 7D lkelihood 
The fiducial $\nu_0$ should reflect the source population so that typical signals spend most of the observation within LISA's most sensitive band, maximizing the accumulated SNR.

% the disadvantage of lag shift
Because the mapping from orbital phase to the TDI response is nonlinear, the best-fit TDI lag yields a slightly biased estimate of $\nu_0$. We therefore use the $7$D likelihood to define a narrow interval for $\nu_0$, followed by an $8$D search that treats $\nu_0$ as a free parameter.
%-------------------

\PRLheading{Nested optimization}
The $\mathcal{F}$-statistic similarly reduces likelihood dimensionality by replacing extrinsic parameters with linear coefficients~\cite{Jaranowski:1998qm}. For a multiharmonic EMRI waveform, however, this decomposition introduces many more coefficients than physical extrinsic parameters. We therefore use the nested optimization developed in Refs.~\cite{Zou:2024jqv,Zou:2024osb}:
\bea
\label{eq:LLR_9D}
L^{\rm 8D}_G = \max\limits_{\theta_{\rm 8D}}\max\limits_{\theta_3} \rho^2(\theta^\prime)\;, \nonumber \\
L^{\rm 7D}_G = \max\limits_{\theta_{\rm 7D}}\max\limits_{n}\max\limits_{\theta_3} \rho^2(\theta^\prime)\;.
\eea
At each PSO evaluation, the Nelder--Mead algorithm~\cite{Nelder-Mead} optimizes $\theta_3$ while $\theta_{\rm 7D}$ or $\theta_{\rm 8D}$ remains fixed. The $7$D likelihood includes an additional lag maximization over $n$, as defined in Eq.~(\ref{eq:lag-shift}).

%-------------------

\endgroup

\begingroup
The outer PSO therefore explores only parameters that reshape the long-term phase and TDI response, while the inner solve profiles over nuisance directions. Unlike a free-coefficient $\mathcal{F}$-statistic, this construction retains the physical correlations of a multiharmonic EMRI waveform. The $7$D stage provides global localization, and the final $8$D stage removes the lag-induced bias in $\nu_0$.
\endgroup

%% ========= 
\PRLheading{Hierarchical search strategy}\label{Hierarchical-Search-Strategy}
Independent local-best PSO runs collect candidate maxima. At each stage, distinct maxima above a threshold $\rho_{\rm threshold}$ define occupied regions in one- and two-dimensional projections. Their coordinate envelope, extended to neighboring bins when the landmark maximum lies near an edge, defines the next prior. The threshold increases as the region contracts but remains low enough to retain a useful population of maxima. Iterations stop when independent runs recover a stable landmark maximum and the region no longer contracts appreciably. Section~\ref{app:method-details} of the Supplemental Material provides the physical motivation, flowchart, peak-region rule, and PSO settings.

%% ========= section result =====
\phantomsection\label{Res}% Results continue without a separate section heading.
We injected two $0.5$-year analytical-kludge signals with target SNR $50$ into stationary Gaussian LISA noise at a $15$-s cadence. The broad priors, spectra, comparison of the $7$D and $8$D searches, and threshold study are provided in Sec.~\ref{app:additional-results} of the Supplemental Material.

\PRLheading{Results}
At each iteration, we retained high-LLR maxima above $\rho_{\rm threshold}$, chosen to preserve enough candidates to define the next contracted region (Sec.~\ref{rho_threshold} of the Supplemental Material). The landmark LLR generally increased as the search region narrowed.

\begin{figure}[H]
	\centering 
	\includegraphics[width=\columnwidth]{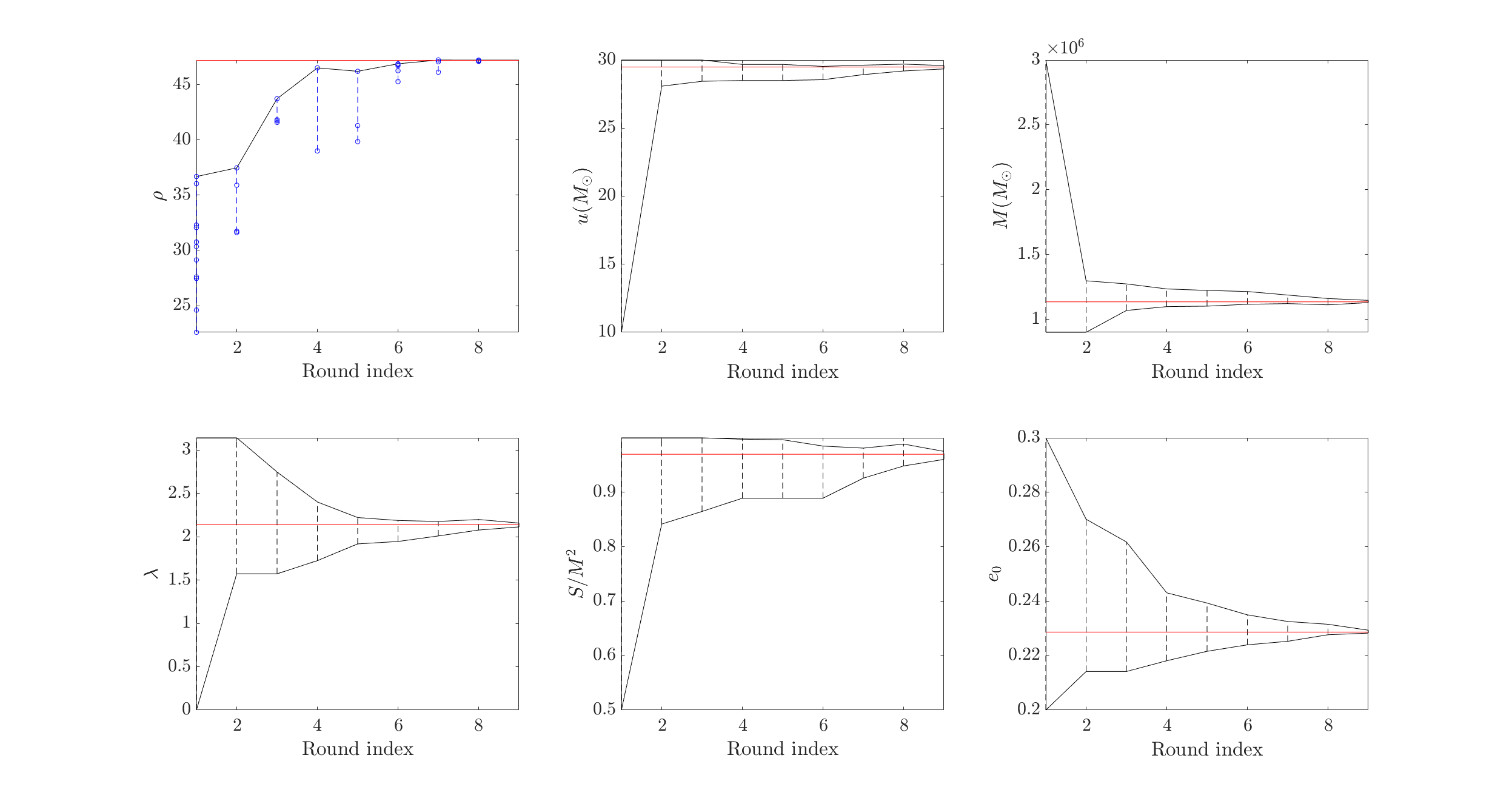}
		\caption{Hierarchical search for injection $1$. Red horizontal lines mark the injected parameters and LLR. Blue circles show landmark LLRs from independent searches; blue dashed lines connect their extrema. Vertical black dashed lines and the enclosing solid black lines mark the contracted ranges.}
	\label{Fig-hie-walker-p1}  
\end{figure}

\begin{figure}[H]
	\centering 
	\includegraphics[width=\columnwidth]{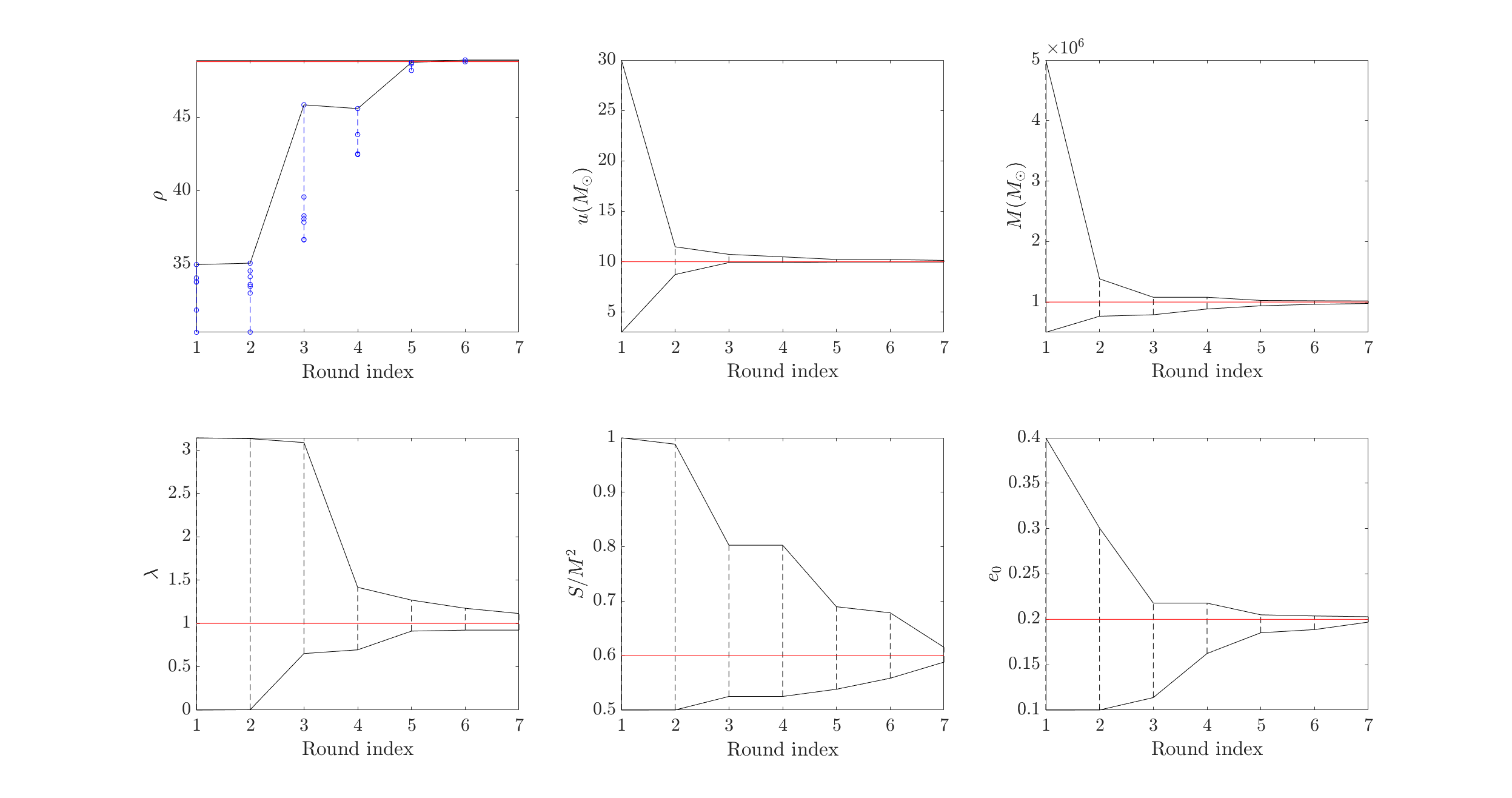}
		\caption{Hierarchical search for injection $2$, with the same conventions as Fig.~\ref{Fig-hie-walker-p1}.}
	\label{Fig-hie-walker-p2}  
\end{figure}

% ------------------ injection 1&2 ------------------
The hierarchy required eight iterations for injection $1$ and six for injection $2$, as shown in Figs.~\ref{Fig-hie-walker-p1} and~\ref{Fig-hie-walker-p2}. In row $1$, column $1$, the landmark LLR increased overall, while its variation across independent PSO runs decreased. Small decreases occurred between iterations $4$ and $5$ for injection $1$ and between iterations $3$ and $4$ for injection $2$. These decreases show that independent runs did not recover the previous landmark maximum at every stage. For injection $2$, the first two landmark LLRs were nearly identical. Row $1$, columns $2$ and $3$ show that the first iteration substantially contracted the $(\mu,M)$ ranges. Subsequent iterations further narrowed these ranges and the other phase-evolution parameters. Further details are provided in Sec.~\ref{peak-pattern} of the Supplemental Material.

Tables~\ref{PE-injection-1} and~\ref{PE-injection-2} report the final estimates from the landmark maximum of the last iteration. Relative errors are defined as $(x_{\rm bestfit}-x_{\rm true})/x_{\rm true}$; sky-location errors are absolute. For the final $8$D estimates, errors in the five phase-evolution parameters other than $\nu_0$ ranged from $2.1\times10^{-4}$ to $2.2\times10^{-2}$. The distance errors were $3.22\%$ and $3.05\%$. Because the $7$D lag estimate is biased, it only defined a several-week interval for the subsequent $8$D search. Treating $\nu_0$ as free then yielded relative errors of $4.3\times10^{-7}$ and $3.5\times10^{-5}$. After accounting for the sky degeneracy, the absolute sky-coordinate errors ranged from $2.7\times10^{-2}$ to $7.2\times10^{-2}$ rad. The recovery differ a lot in $\theta_3$. The reconstructed TDI channels achieved $\mathrm{ff}_{\rm AE}=0.989$ and $0.971$. 

% hierarchical injection, rough look

% PE  injection 1
\begin{table}[H]
	\centering
	\scriptsize
	\setlength{\tabcolsep}{1.5pt}
	\resizebox{\columnwidth}{!}{%
	\begin{tabular}{@{}cccc@{}}
		\hline\hline
		Parameter   &Injection  &\makecell{Best-fit and error \\ 7D} &\makecell{Best-fit and error \\ 8D}\\
		\hline
		$\mu(M_\odot)$  &$29.49$   &\fitresult{29.44702}{(-1.457587\times10^{-3})}   &\fitresult{29.44899}{(-1.390558\times10^{-3})} \\
		$M(10^6M_\odot)$    &$1.134945$   &\fitresult{1.136182}{(1.089757\times10^{-3})}   &\fitresult{1.138361}{(3.009658\times10^{-3})} \\
		$\lambda$(rad)  &$2.1422$    &\fitresult{2.138088}{(1.919588\times10^{-3})}    &\fitresult{2.133349}{(-4.131734\times10^{-3})} \\
		$S/M^2$ &$0.9697$   &\fitresult{0.9683166}{(-1.42664\times10^{-3})}   &\fitresult{0.966788}{(-3.002991\times10^{-3})} \\
		$e_0$ &$0.2286567$    &\fitresult{0.22959}{(4.081959\times10^{-3})}  &\fitresult{0.228705}{(2.114428\times10^{-4})} \\
		\hline
		$\nu_0$(mHz) &$0.7380463$     &$\sim$  &\fitresult{0.738046}{(-4.255495\times10^{-7})} \\
		\hline
		$\theta_s$(rad) &$0.4989445$     &\fitresult{0.4705113}{(-2.843318\times10^{-2})}  &\fitresult{2.674157}{(-3.150884\times10^{-2})^{*}} \\
		$\phi_s$(rad)   &$2.232797$     &\fitresult{2.169612}{(-6.318524\times10^{-2})}  &\fitresult{5.347037}{(-2.735263\times10^{-2})^{*}} \\
		\hline
		$\theta_k$(rad) &$1.5221$     &$\sim$   &\fitresulttwo{0.567697} \\
		$\phi_k$(rad) &$3.9467$     &$\sim$  &\fitresulttwo{-0.760648} \\
		$\phi_0$ &$2.041532$     &$\sim$  &\fitresulttwo{2.125061} \\
		$\tilde{\gamma}_0$ &$5.659686$     &$\sim$   &\fitresulttwo{0.797133} \\
		$\alpha_0$ &$1.175479$     &$\sim$   &\fitresulttwo{-0.137371} \\				
		\hline
		$D_L$(Gpc) &$1.5004$  &$\sim$  &\fitresult{1.548678}{(0.0322)} \\
        \hline
		$\rho$ &\makecell{$47.199341$($7$D)\\$47.532608$($8$D)}  &$47.648995$  &$47.670344$  \\
        \hline
		$\mathrm{ff}_{\rm AE}$ &$1$  &$\sim$  &$0.988949$ \\
		\hline\hline
	\end{tabular}}
		\caption{Parameter estimation for injection $1$. The fitting factor $\mathrm{ff}_{\rm AE}$ is defined in Eq.~(\ref{Eq:ff}); the superscript `*' denotes the degenerate sky location $(\pi-\theta_s,\pi+\phi_s)$.}\label{PE-injection-1}
\end{table}

% hierarchical injection, rough look

%  PE  injection-2
\begin{table}[H]
	\centering
	\scriptsize
	\setlength{\tabcolsep}{1.5pt}
	\resizebox{\columnwidth}{!}{%
	\begin{tabular}{@{}cccc@{}}
		\hline\hline
		Parameter   &Injection  &\makecell{Best-fit and error \\ 7D} &\makecell{Best-fit and error \\ 8D}\\
		\hline
		$\mu(M_\odot)$  &$10.0$    &\fitresult{10.01383}{(1.382600\times10^{-3})}   &\fitresult{10.04678}{(4.678300\times10^{-3})} \\
		$M(10^6M_\odot)$    &$1.0$    &\fitresult{0.9972095}{(-2.790524\times10^{-3})}   &\fitresult{0.9966528}{(-3.347165\times10^{-3})} \\
		$\lambda$(rad) &$1.0$    &\fitresult{1.013862}{(1.386200\times10^{-2})}   &\fitresult{1.021875}{(2.187500\times10^{-2})} \\
		$S/M^2$ &$0.6$    &\fitresult{0.6017690}{(2.948333\times10^{-3})}   &\fitresult{0.602919}{(4.865000\times10^{-3})} \\
		$e_0$ &$0.2$    &\fitresult{0.1998850}{(-5.750000\times10^{-4})}   &\fitresult{0.1973040}{(-1.348000\times10^{-2})} \\
		\hline
		$\nu_0$(mHz) &$1.0$    &$\sim$   &\fitresult{1.000035}{(3.500000\times10^{-5})} \\
		\hline
		$\theta_s$(rad) &$1.0$   &\fitresult{2.137027}{(4.565654\times10^{-3})^{*}}   &\fitresult{2.069450}{(7.214265\times10^{-2})^{*}} \\
		$\phi_s$(rad)   &$1.0$    &\fitresult{4.108400}{(-3.319265\times10^{-2})^{*}}   &\fitresult{4.081102}{(-6.049065\times10^{-2})^{*}} \\
		$\theta_k$(rad) &$1.0$     &$\sim$  &\fitresulttwo{0.574012} \\
		$\phi_k$(rad) &$1.0$     &$\sim$  &\fitresulttwo{1.514739}\\
		$\phi_0$(rad) &$0.0$     &$\sim$  &\fitresulttwo{-0.496054} \\
		$\tilde{\gamma}_0$(rad) &$0.0$     &$\sim$   &\fitresulttwo{1.881838} \\
		$\alpha_0$(rad) &$0.0$     &$\sim$  &\fitresulttwo{1.289012} \\
	    \hline
		$D_L$(Gpc) &$0.5309$  &$\sim$  &\fitresult{0.547118}{(0.0305)} \\
        \hline
		$\rho$ &\makecell{$48.789520$($7$D)\\$48.743604$($8$D)}  &$48.896010$  &$48.799397$  \\
        \hline
		$\mathrm{ff}_{\rm AE}$ &$1$  &$\sim$  &$0.971143$ \\
		\hline\hline
	\end{tabular}}
		\caption{Parameter estimation for injection $2$, with the same conventions as Table~\ref{PE-injection-1}.}\label{PE-injection-2}
\end{table}

%#############################

\PRLheading{Discussion}\label{Discussion}
The central obstacle to coherent EMRI detection is not evaluating the likelihood but navigating its high-dimensional, strongly multimodal landscape. We show that higher-likelihood secondary maxima concentrate progressively around the global maximum, a structure arising from partial harmonic matches in the multiharmonic waveform. By exploiting this pattern through reduced-dimensional profile likelihoods—seven-dimensional for global localization and eight-dimensional for refinement—combined with nested optimization and particle-swarm peak collection, we turn secondary maxima into guides for adaptive prior contraction. 

For the two generic, spinning EMRI injections, high-LLR secondary maxima became increasingly concentrated around the primary maximum. This behavior appears in the partial-harmonic analysis (Figs.~\ref{fig:harmonics} and~\ref{fig:peak_landscape_1} in the Supplemental Material) and the $7$D hierarchical searches (Figs.~\ref{Fig-hie-walker-p1} and~\ref{Fig-hie-walker-p2}). Additional projections are shown in Figs.~\ref{Fig-hie-p1} and~\ref{Fig-hie-p2} of the Supplemental Material. This structure provides the practical basis for adaptive contraction. Starting from astrophysically broad priors, the hierarchy localized both signals, and the final $8$D searches recovered the complete $14$-parameter EMRI signals. Together, these results demonstrate an end-to-end coherent search for generic EMRIs over astrophysically broad priors.

% 2026/08/05
Our work demonstrates a coherent hierarchical strategy that turns the multimodal likelihood structure of generic EMRIs into a search asset rather than an obstacle, and provide a practical solution to coherent end-to-end searches over broad priors. Extending this framework to more realistic noise and to more accurate waveform templates will ready the pipeline for application to upcoming space-borne detector data.
%Our work offers a preliminary solution to the coherent end-to-end search for generic EMRI signals over broad priors. Future work will focus on the pipeline %Future work will focus on selecting a population-informed fiducial $\nu_0$ and incorporating more accurate waveforms generated with FastEMRIWaveforms (FEW)~\cite{Katz:2021yft}.
%#############################

\begin{acknowledgments}
We gratefully acknowledge financial support from the National Key R\&D Program of China (Grant No.~2021YFC2203002), the National Science and Technology Major Project of China (No. 2024ZD1100601) and the National Natural Science Foundation of China (Grants No.~12173071 and No.~12473075). The numerical computations were carried out on the high-performance computing cluster at the School of Fundamental Physics and Mathematical Sciences, Hangzhou Institute for Advanced Study, UCAS. We thank RunQiu Liu, Shutong Liu, Yan Wang, Qun-ying Xie, Pin Shen, Qian-yun Yun, Runming Yao, Xue-hao Zhang, and Shao-dong Zhao for discussions and technical assistance.
\end{acknowledgments}

%\clearpage

\clearpage
\section*{\NoCaseChange{Supplemental Material}}
\appendix
\renewcommand{\appendixname}{}
\setcounter{secnumdepth}{3}

\section{Model}
\label{model}
In this section, we depict the TDI~\cite{Tinto:2004wu} that suppress the laser frequency noise, the power spectral density (PSD) of the LISA noise model adopt from the LDC manual~\cite{LDC-code-maunal}. The analytical kluge (AK)~\cite{Barack:2003fp} waveform is also briefly introduced. All are consistent with LDCs and our previous series works~\cite{Zou:2024jqv,Zou:2024osb}.
%-------------------
\subsection{TDI}
The space-based gravitational-wave detector consists of three spacecraft arranged in a triangular formation. Each laser link connects two spacecraft, identified by the triplet $\{s,l,r\}$, where $s$, $l$, and $r$ denote the sender, the link, and the receiver, respectively. The unit vector along arm $l$, represented as $\widehat{n}_l$,  is obtained from the orbital trajectories of the detector. For a GW with plus/cross polarizations $h_{+,\times}$, polarization tensors $\epsilon^{+,\times}$, and polarization angle $\psi$, the strain projected onto $\widehat{n}_l$, written as $H_l$, takes the form:
  \bea
  \label{eq:H_l}
    H_l =& F_l^{+}h_{+}+F_l^{\times}h_{\times}\;,   \nonumber \\
  \begin{bmatrix}F_l^+ \\ F_l^{\times} \end{bmatrix}=&\begin{bmatrix} \cos(2\psi) & -\sin(2\psi)\\ \sin(2\psi)& \cos(2\psi)\end{bmatrix}\begin{bmatrix} (\widehat{n}_l \otimes \widehat{n}_l):\epsilon^{+}\\(\widehat{n}_l \otimes \widehat{n}_l):\epsilon^{\times}\\ \end{bmatrix}\;,  \\   
  \nonumber 
  \eea
 where $F_l^{+, \times}$ represent the antenna pattern functions for link $l$, which encode the directional sensitivity to incoming GWs. The contraction operation $:$ is defined as $U:V = \sum_{i,j} U_{ij}V_{ij}$ for any tensors $U$ and $V$, while the outer product $\otimes$  yields $(a \otimes b)_{ij}=a_i b_j$ for vectors $a$ and $b$. 

 The GW-induced strain response of a single link $l$, denoted $y^{\rm GW}_{slr}$, is defined as the difference between the $H_l$ values evaluated at the sender and receiver:
 \bea
 y^{\rm GW}_{slr}(t) =& \frac{H_l(t-\hat{k}\cdot\widehat{R}_s-L_l) - H_l(t-\hat{k}\cdot\widehat{R}_r)}{2(1-\hat{k}\cdot\widehat{n}_l)}\;.
 \label{eq:TDI_y_slr}
 \eea 

 The expression for $y^{\rm GW}_{slr}$ involves time delays and projections determined by the GW propagation direction $\hat{k}$, the detector's orbital geometry (specifically, the spacecraft positions $\widehat{R}_s$ and $\widehat{R}_r$), the unit arm vector $\widehat{n}_l$, and the arm length $L_l$. The index $l$ follows the cyclic ordering $1 \to 2 \to3 \to 1$ of the triplets $\{s,l,r\}$; a positive sign is associated with this cyclic order, and a negative sign otherwise.

 We construct the Michelson-type TDI variables X, Y, and Z following the first-generation TDI scheme used in the LDCs for EMRI analyses:
\begin{align}
 X = &-2i\sin(\omega L)*\nonumber \\ & \left[e^{-i2\omega L}(y_{1-32}-y_{123}) +  e^{-i\omega L}(y_{231}-y_{3-21}) \right]\;,  \nonumber \\
 Y=&-2i\sin(\omega L)* \nonumber \\ & \left[e^{-i2\omega L}(y_{2-13}-y_{231}) +  e^{-i\omega L}(y_{312}-y_{1-32}) \right]\;, \nonumber \\
 Z=&-2i\sin(\omega L)* \nonumber \\ & \left[e^{-i2\omega L}(y_{3-21}-y_{312}) + e^{-i\omega L}(y_{123}-y_{2-13}) \right]\;. \label{eq:TDI-1} \\ \nonumber  
\end{align}

In the first-generation TDI formulation, all arm lengths $L_l$ are assumed to be equal to a constant $L$, and the GW frequency $\omega$ is taken as the time derivative of the GW phase: $\omega=\frac{\partial \Phi^{\rm GW}}{\partial t}$. From the Michelson channels, we obtain the noise-orthogonal combinations $A, E$, and $T$ through the following transformation:
	\bea
A =& \frac{Z-X}{\sqrt{2}}\;, \nonumber \\ 
E =& \frac{X-2Y+Z}{\sqrt{6}}\;, \nonumber\\ 
T =& \frac{X+Y+Z}{\sqrt{3}}\;. \label{eq:TDI-2}\\ 
\nonumber
\eea
As the $T$ channel is largely insensitive to GWs, we restrict our analysis to the $A$ and $E$ channels, which are computed by substituting the single-link strain response $y^{\rm GW}_{slr}$ into Eq.~(\ref{eq:TDI-1}) and (\ref{eq:TDI-2}).

%The data model is 
%\bea
%\overline{d}^I = \overline{h}^I + \overline{n}^I\;,
%\eea
%-------------------
\subsection{PSD}
\label{psd}
The theoretical PSD model for LISA employed in this work follows the formulation given in~\cite{LDC-code-maunal}. Because the $A$ and $E$ channels exhibit identical power spectral densities, we introduce a common notation $S_n(f)$ for their noise spectra in the following analysis:
\begin{align}
\label{eq:PSD}
S_n(f)&=8\sin^2 x\big\{4(1+\cos x+\cos^2x)S^{\rm Acc}\nonumber\\
&\hspace{20mm} +(2+\cos x)S^{\rm IMS}\big\},\qquad x=\omega L\;.
\end{align}

In the above expression, $f$ is the Fourier frequency, $\omega = 2\pi f$ denotes the angular frequency, and $L$ stands for the detector arm length, which is treated as a constant under the first-generation TDI approximation. Based on the LISA reference noise model "SciRDv1"~\cite{LDC-code-maunal}, the acceleration noise $S^{\rm Acc}$ and the instrumentation optical metrology system noise $S^{\rm IMS}$ are specified as:
\bea
%\begin{align}
	S^{\rm Acc}(f)=& \frac{9.0\times10^{-30}}{(2\pi fc)^2}\big[1+\big(\frac{0.4 {\rm mHz}}{f}\big)^2\big]\big[1+\big(\frac{f}{8{\rm mHz}}\big)^4\big]\frac{1}{\text{Hz}}\;,  \nonumber \\ 
	S^{\rm IMS}(f)=&2.25\times 10^{-22}\big(\frac{2\pi f}{c}\big)^2\big[1+(\frac{2 {\rm mHz}}{f})^4\big]\frac{1}{\text{Hz}}\;, \nonumber \\  
%\end{align}
\label{eq:LDC_noise_model} 
\eea
where $c$ is the speed of light.

Given the PSD, the noise-weighted inner product between two signals  $\overline{a}$
and $\overline{b}$ is defined as follows:
\bea
(\overline{a}|\overline{b}) = \frac{1}{Nf_s}\sum_{k=0}^{N-1}\frac{\widetilde{a}_k\widetilde{b}^\ast_k + \widetilde{a}^\ast_k\widetilde{b}_k}{S_n(f_k)}\;,
\label{eq:innerproduct}  
\eea
where $\widetilde{a}$ represents the discrete Fourier transform (DFT) of the time-domain signal $\overline{a}$, and $\widetilde{a}^{\ast}$ is its complex conjugate (the same notation applies to $\overline{b}$). The spectral resolution of the DFT is determined by the sampling frequency $f_s$, the signal length $N$, and the discrete frequency bins $f_k = kf_s/N$ for $k = 0,1,\ldots,N-1$.

For the noise-orthogonal TDI channels $\{A,E\}$, the combined signal-to-noise ratio (SNR) for signal $\overline{a}$, and the fitting factor between signal $\overline{a}$ and $\overline{b}$ are obtained from the inner-product structure as follows:
\bea
\label{Eq:ff}
{\rm SNR}^2_a =& (\overline{a}^A|\overline{a}^A)+(\overline{a}^E|\overline{a}^E)\;, \nonumber \\ 
{\rm ff}_{\rm AE} = & \frac{(\overline{a}^A|\overline{b}^A)+(\overline{a}^E|\overline{b}^E)}{\sqrt{(\overline{a}^A|\overline{a}^A)+(\overline{a}^E|\overline{a}^E)}\sqrt{(\overline{b}^A|\overline{b}^A)+(\overline{b}^E|\overline{b}^E)}}\;,\\ \nonumber
\eea 
where the component in one channel vanishes when evaluating the corresponding quantity for an individual signal (analogous SNR notation holds for $\overline{b}$).
%-------------------
\subsection{Waveform}
\label{waveform}
Accurately modeling EMRI waveforms is challenging and computationally demanding when employing black hole perturbation theory~\cite{Jiang:2025mna} or self-force methods~\cite{Barack:2018yvs}, which are required to account for relativistic orbital evolution and radiation reaction. For data analysis purposes, the phenomenological "kluge" waveform has been developed to leverage its advantage of rapid computation. Although less physically accurate, it retains the key characteristics\textemdash such as pericenter precession and Lense-Thirring precession modeled via post-Newtonian (PN) descriptions\textemdash that produce a signal morphology similar to that of true EMRI signals~\cite{Barack:2003fp}. Consequently, the AK waveform has been adopted in the MLDCs, LDCs, and TDCs. While the more recent FEW model~\cite{Katz:2021yft}, which combines accuracy with speed, is becoming increasingly popular for data analysis, we employ the AK waveform in this work to maintain consistency with the current data challenges and our previous studies. We note that the methodology developed for AK is directly applicable to FEW, requiring primarily an adaptation to address certain computational specificities.

The orbital equations addressing the orbital motion, pericenter precession and Lense-Thirring precession described by PN terms are given in Eq.~(27-31) in the original AK paper~\cite{Barack:2003fp},which is solved using Runge-Kutta methods. The full parameter set $\Theta$ consist of three subsets as following,
\bea
\Theta = \theta_1 \cup \theta_2  \cup \theta_3\;,
\eea
with different features summarized in Tab.~\ref{tab-1-reduced-diemsnionality}. 
\begin{table}[htb]
	\centering
	\resizebox{\columnwidth}{!}{%
	\begin{tabular}{cc}
		\hline\hline
		Parameter Set   &Description\\
		\hline
		$\theta_1 = \{\mu,M,\lambda,S/M^2,e_0,\nu_0\}$
		& \makecell{Time dependent \\on the orbital evolution,\\ GW phase-coupled} \\ \hline
		$\theta_2 = \{\theta_s, \phi_s\}$ & \makecell{Time dependent \\on the TDI delay \\due to detector motion}  \\ \hline
		$\theta_3 = \{\theta_k, \phi_k, \phi_0,\tilde{\gamma}_0, \alpha_0,D\}$ & \makecell{Time independent,\\ modulating the time independent \\ amplitude and initial phases} \\		 
		\hline\hline
	\end{tabular}}
	\caption{Illustration of the three parameter sets in $\Theta$.
	}\label{tab-1-reduced-diemsnionality}
\end{table}

\begin{figure}[htbp]
	\centering
	\includegraphics[width=0.45\textwidth]{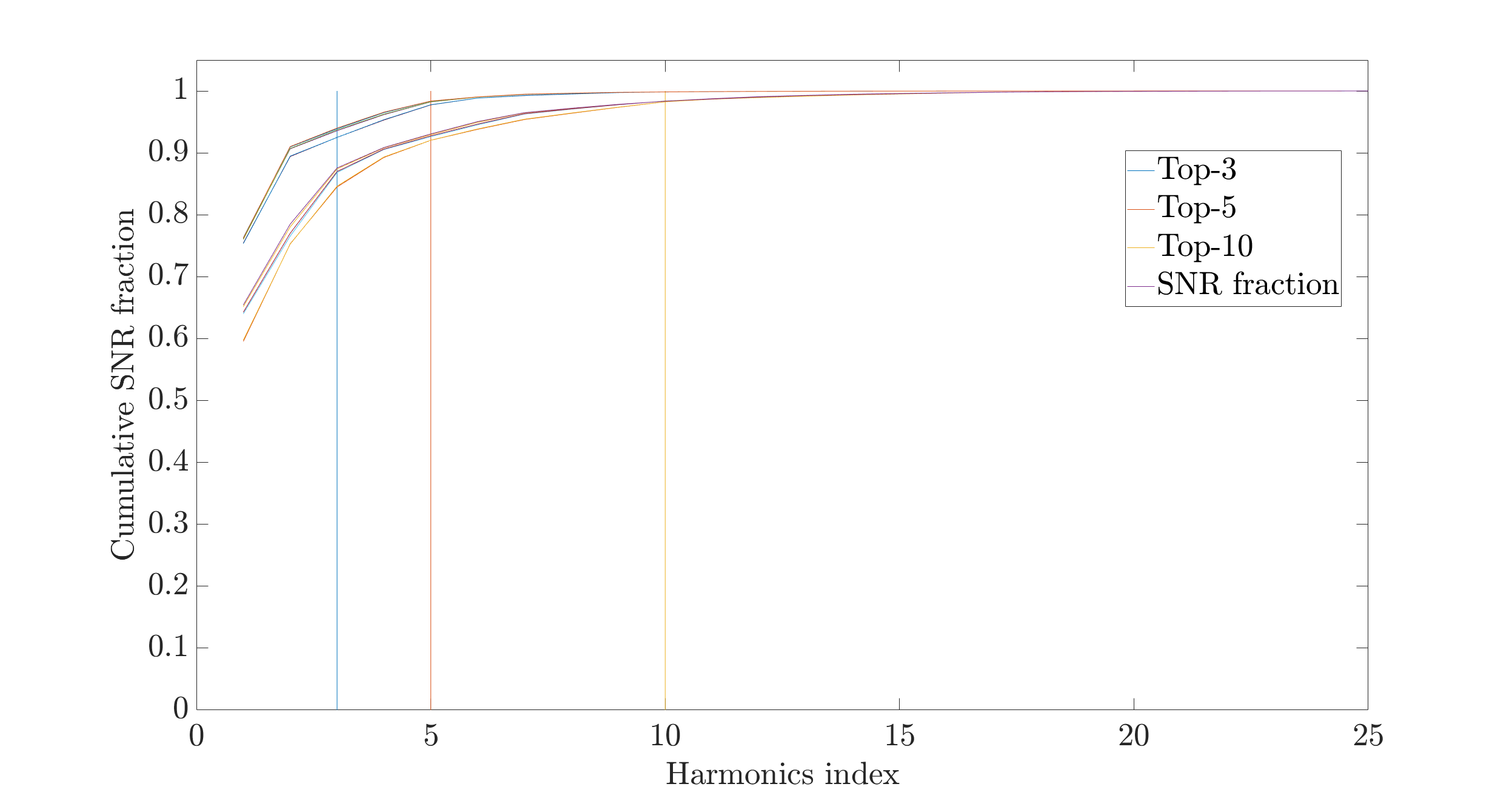}
	\caption{Illustration of the cumulative SNR for $25$ harmonics ($n=1,2,3,4,5, l=2, m=-2,-1,0,1,2$) of AK waveform in descending order for $12$ source parameters with different mass ratio, eccentricity and MBH's spin.}
	\label{fig:harmonics}
\end{figure}

Regarding eccentricity and precession, the EMRI polarized waveform $\overline{h}(\Theta)_{+,\times}$ is a superposition of multiple harmonics as follows:
\bea
   \overline{h}(\Theta)_{+,\times} =& \sum_{\rm nlm} A_{\rm nlm}(e(t),\nu(t))e^{i(n\phi(t)+l\tilde{\gamma}(t) +m\alpha(t))}\;.\nonumber  \\ 
\eea
As shown in Fig.~\ref{fig:harmonics}, the top-$10$ dominant harmonics typically account for more than $95\%$ of the total SNR for most EMRI sources. Therefore, to reduce computational cost while maintaining an acceptable SNR loss, we select the top-$10$ harmonics with $n=2,3$ for the EMRI injections in this paper. Further discussion can be found in ~\cite{Zou:2024jqv,Zou:2024osb}.

%############################## sec 2 ##############################
\section{Additional Methodological Details}\label{app:method-details}
\subsection{Particle Swarm Optimization}
PSO is a point-based global optimization method belonging to the evolutionary algorithm family. It employs $N_p$ particles to explore and exploit the parameter space of a high-dimensional, multimodal fitness function  $f(\overline{x})$ over $N_{\rm iter}$ iterations, aiming to locate the global optimum $\overline{x}_{*}$, such that $f(\overline{x}_{*})>f(\overline{x}), \forall \overline{x}$. In particular, with $N_{\rm run}$ independent PSO searches, the probability that at least one search successfully locates the global optimum approaches unity, given by $1-p_0^{N_{\rm run}}$, where  $p_0$ is the probability that a single PSO search fails. In our previous works~\cite{Zou:2024jqv,Zou:2024osb}, $N_p=40$, $N_{\rm iter}=10000$, and $N_{\rm run}\ge 6$ exhibit satisfactory performance in searching for EMRI signals.

The velocity $\overline{v}$ and position $\overline{x}$ are updated according to the following rules:
\begin{align}
\overline{v}_{i}(t+1)={}&w\overline{v}_{i}(t)
 +c_1r_1[\overline{p}_{i}(t)-\overline{x}_{i}(t)]\nonumber\\
 &+c_2r_2[\overline{g}(t)-\overline{x}_{i}(t)]\;,
\label{eq:velocity_update}\\
\overline{x}_{i}(t+1)={}&\overline{x}_{i}(t)+\overline{v}_{i}(t+1)\;.
\label{eq:position_update}
\end{align}
where the three terms in Eq.~(\ref{eq:velocity_update}) account for the inertia of the particle itself, as well as the local and global influences from other particles. Here $i$ indexes the particle from $1$ to $N_p$, and $t$ denotes the iteration number. $\overline{p}_{i}$ represents the personal best position of the $i$-th particle, and $\overline{g}$ denotes the global best position across all particles. Both gradually approach the global optimum $\overline{x}_{*}$ as the search converges. $r_1$ and $r_2$ are random numbers uniformly distributed in $\left[0, 1\right]$, introduced to provide stochasticity. The common values are $c_1=c_2=2$, the inertia weight $w$ linearly decreasing from $0.9$ to $0.4$, $V_{\rm max}=0.5$ and a clamping operation is applied to the velocities, ensuring  $|v_i^j| \le V_{\rm max}$.

To enhance the exploration capability of PSO, the local-best PSO variant~\cite{Kennedy:2007} is usually applied. In this approach, a representative local best position $\overline{p}_{\text{local}, i}$
is used in place of the global best $\overline{g}$. This local best is selected from a ring topology for each particle as follows:
\bea
f(\overline{p}_{\text{local}, i}(t)) = \max\limits_{j\in \mathcal{N}_i} f(\overline{p}_j(t))\;,
\eea
where $\mathcal{N}_i=\{i-1, i, i+1\}, i \in 1,2,3,...,N_p$ and the first and last particle are circularly connected.
%-------------------
\subsection{Full Hierarchical-Search Construction}
\label{app:hierarchy-details}
\begin{figure}[!t]
	\centering
	\includegraphics[width=\columnwidth]{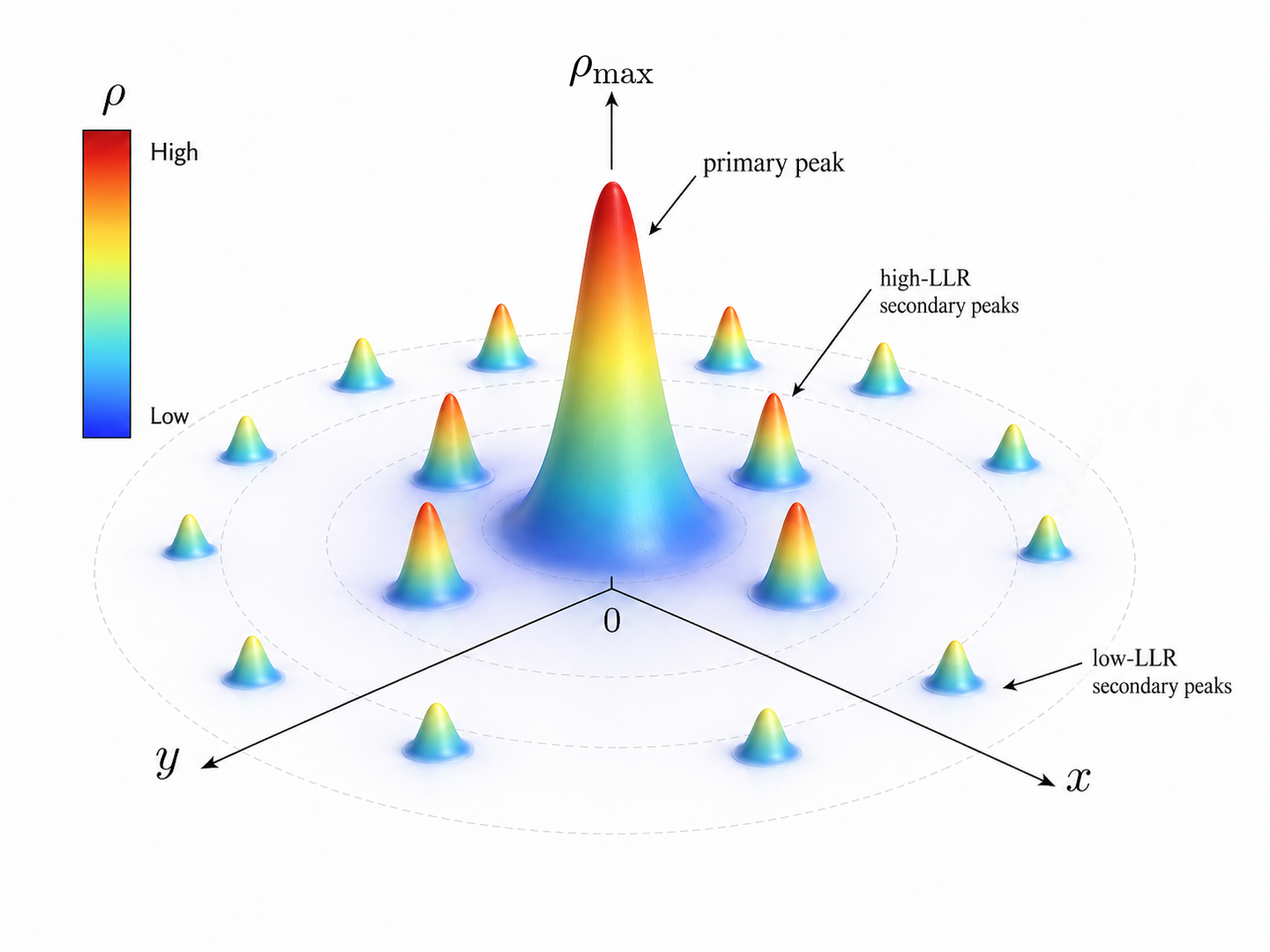}
	\caption{Conceptual illustration of the expected peak landscape in likelihood surface, $\rho$ is defined in Eq.~(\ref{eq:8D-LLR}). Note this figure is produced by artificial intelligence tool Gemini.}
	\label{fig:peak_landscape_1}
\end{figure}

Considering that EMRI waveforms comprise multiple harmonics (e.g., $25$ in the LDC‑$1.2$ Radler dataset and more in realistic scenarios), each with a distinct SNR as shown in Fig.~\ref{fig:harmonics}, the low‑SNR harmonics are easily submerged in the noise background and thus missed during matched filtering. The more low‑SNR harmonics that are missed, the lower the remaining SNR becomes, and consequently the greater the deviation of local extrema (secondary peaks) on the likelihood surface from the global maximum (primary peak). Based on this property, a general global characteristic of the distribution of secondary peaks relative to the primary peak can be summarized: \textbf{high‑LLR secondary peaks tend to cluster increasingly tightly around the primary peak}, as conceptually illustrated in Fig.~\ref{fig:peak_landscape_1}. Note the LLR defined in Eq.~(\ref{eq:8D-LLR}) is the matched filtering SNR.

\begin{figure}[!t]
	\centering
	\includegraphics[width=\columnwidth]{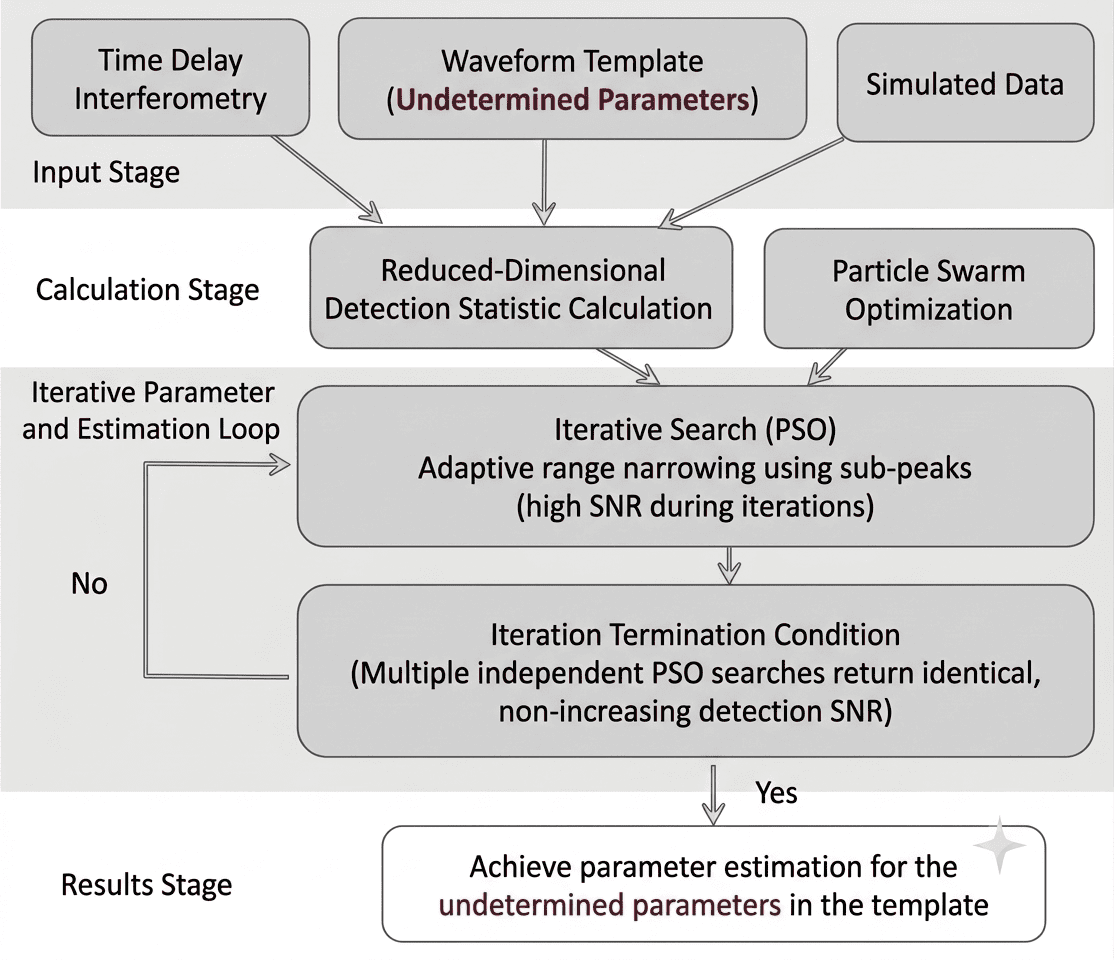}
	\caption{The flowchart of the hierarchical search, also produced by Gemini.}
	\label{fig:flowchart-hierarical-search}
\end{figure}

Given the above general feature of secondary-peaks versus primary-peak distribution, a natural staged iterative search strategy is as follows: \textbf{starting from a large prior range, collect high-LLR secondary peaks, use their clustering property to estimate a smaller search range, and iterate this process until the maximum LLR returned by multiple independent PSO searches stabilizes and no longer increases}. A flowchart of this procedure is provided in Fig.~\ref{fig:flowchart-hierarical-search}.

\clearpage

The key control parameter of this strategy is the LLR threshold selected at each iteration. This threshold must ensure that a sufficient number of high-LLR secondary peaks are collected so that their tight clustering behavior becomes statistically significant. Consequently, the estimated smaller search range is more likely to enclose the primary peak, thereby facilitating its identification.

\begin{figure}[htbp]
	\centering
	\includegraphics[width=0.45\textwidth]{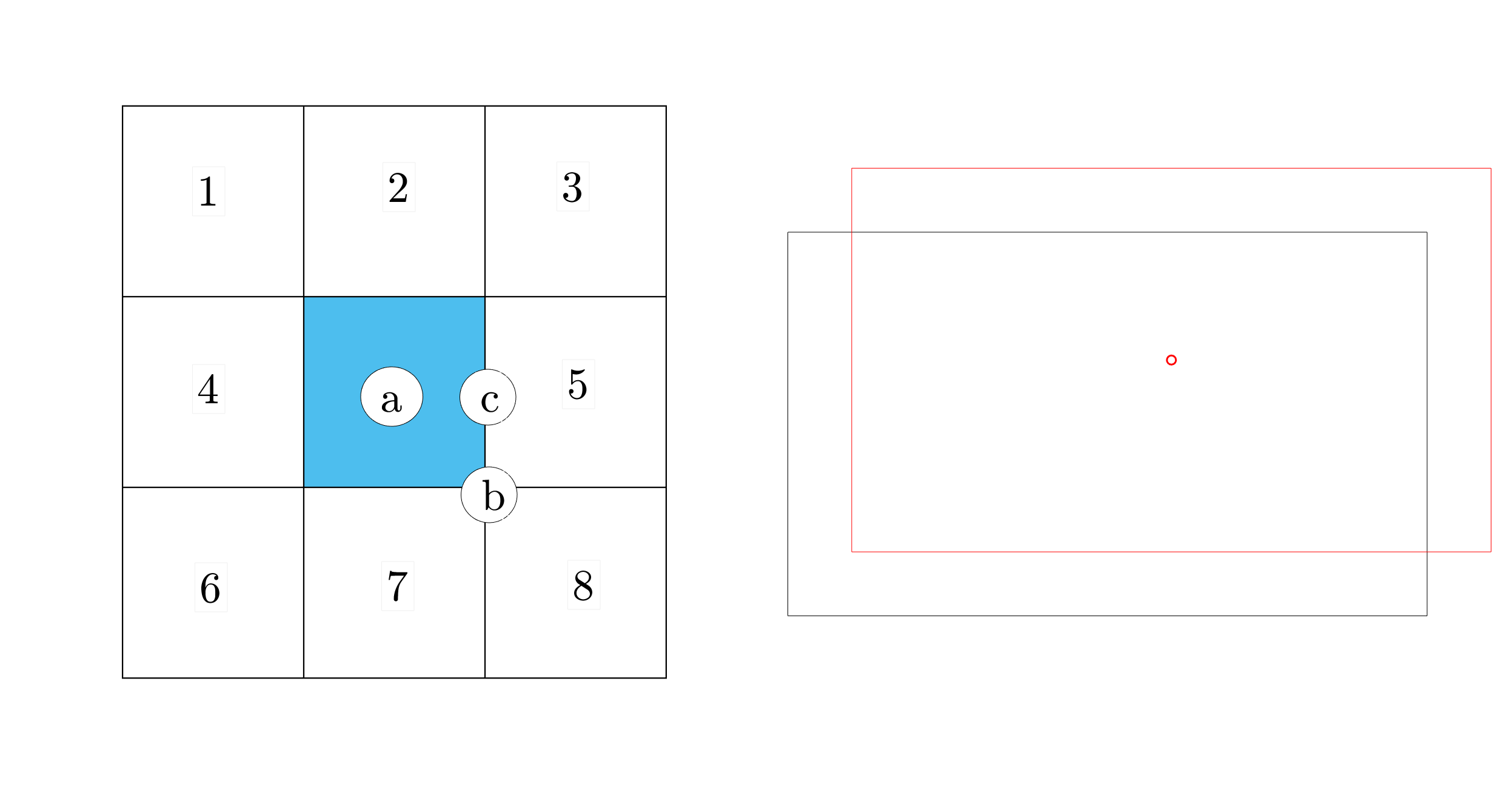}
	\caption{Additional strategy in the hierarchical search.}
	\label{Fig:tricks-hierarchical-search}
\end{figure}

However, due to limited computational resources, we may sometimes obtain only a small number of high-LLR secondary peaks. In such cases, we still need to apply certain heuristic strategies to carry out the hierarchical search. As illustrated in Fig.~\ref{Fig:tricks-hierarchical-search}, the right panel shows that when the number of high-LLR peaks is insufficient, the identified peak region (shaded blue) may fail to contain the primary peak; nevertheless, extending this region together with its eight neighboring regions (for each $2$D slice) can still enclose the primary peak. Sometimes, extending only certain neighboring regions is sufficient. For example:
\begin{enumerate}
	\item extend all eight neighboring regions if the landmark peak is located at position `a';
	\item extend the neighboring regions labeled $\{5,7,8\}$ or simply  $\{8\}$ if the landmark peak is located at position `b'; 
	\item  extend the neighboring regions labeled  $\{2,3,5,7,8\}$ if the landmark peak is located at position `c'.
\end{enumerate} 
The right panel demonstrates that the landmark secondary peak (i.e., the one with the highest LLR) moves closer to the primary peak as the hierarchical search progresses. Therefore, shifting the peak region to be centered on this landmark peak (red circle) helps improve the estimation of the smaller search range, especially in the later stages of the hierarchical search.
%\begin{figure}[htbp]
%	\centering 
%	\includegraphics[width=0.23\textwidth]{1116_peak_region_neighbour_1.png}
%	%\hfill
%	%\par  % 强制换行
%	%\hspace{-2.5mm} 
%	\includegraphics[width=0.23\textwidth]{1116_peak_region_neighbour_2.png}
%	%\hspace{-7.5mm} 
%	\caption{Additional strategy in the hierarchical search.}
%	\label{Fig:tricks-hierarchical-search}  
%\end{figure}
%================ sec 3
\section{Computational Cost}\label{app:cost}
The hierarchical search pipeline is implemented using \texttt{OpenMP} and \texttt{OpenMPI}. The former is employed to parallelize internal likelihood computations, e.g., the polarized waveform calculation for each harmonic and the multiple single-arm response calculations in TDI. The latter is used to parallelize external likelihood evaluations, e.g., the likelihood evaluation of $N_p$ particles. Due to limitations in the computational resources available, multiple independent PSO searches must be performed serially. The computational cost of each component and the relevant cluster information are provided in Tab.~\ref{Tab:cost}.
	\begin{table}[htb]
	\centering
	\begin{tabular}{c|c}
		\hline
	    Single Likelihood evaluation cost (8D) &$\sim0.4$ seconds \\
		Individual PSO search ($N_{\rm iter}=10000$, $8$D) &$\sim1$ hour \\
		Single Likelihood evaluation cost (7D) &$\sim8$ seconds \\
        Individual PSO search ($N_{\rm iter}=10000$, $7$D) &$\sim24$ hours \\
    \end{tabular}   
    \begin{tabular}{c|c}    
        \hline
        \makecell{Hierarchical search using $7$D}  & \makecell{$\sim 6$ independent PSO \\ searches  for each iteration, \\ $6\sim8$ iterations, \\ total $\sim 50$ independent \\ PSO searches} \\
        \hline 
        Hierarchical search cost & \textbf{$\sim 2$ months} \\
		\hline
		Number of threads (OpenMP) & $64$ \\
		Number of nodes (OpenMPI) & $40$   \\
		CPU memory & $200$ GB       \\
		CPU Clock Rate & $1.5$ GHz    \\
		Total CPU hours & \textbf{$\sim3\times 10^6$}   \\
		\hline
	\end{tabular}
	\caption{Demonstration of the computational cost for hierarchical searching a duration of $0.5$-year data. }\label{Tab:cost}
\end{table}
Note that the current CPU (Central Processing Unit) pipeline is not sufficiently efficient: a full hierarchical search for $0.5$-year data from start to finish requires at least $2$ months. The computational cost will increase further when the data duration extends to several years, comparable to the operational lifetime of LISA. There is a pressing need to introduce graphics processing unit (GPU) acceleration, which is expected to accelerate the hierarchical search by a factor of up to $20$, thereby reducing the computational cost to within a week. This constitutes part of our ongoing work.
%############################## sec 
\section{Additional Results}\label{app:additional-results} 
% -------------------------------------
\subsection{Injection and search range} 
\begin{figure}[htbp]
	\centering 
	\includegraphics[width=0.45\textwidth]{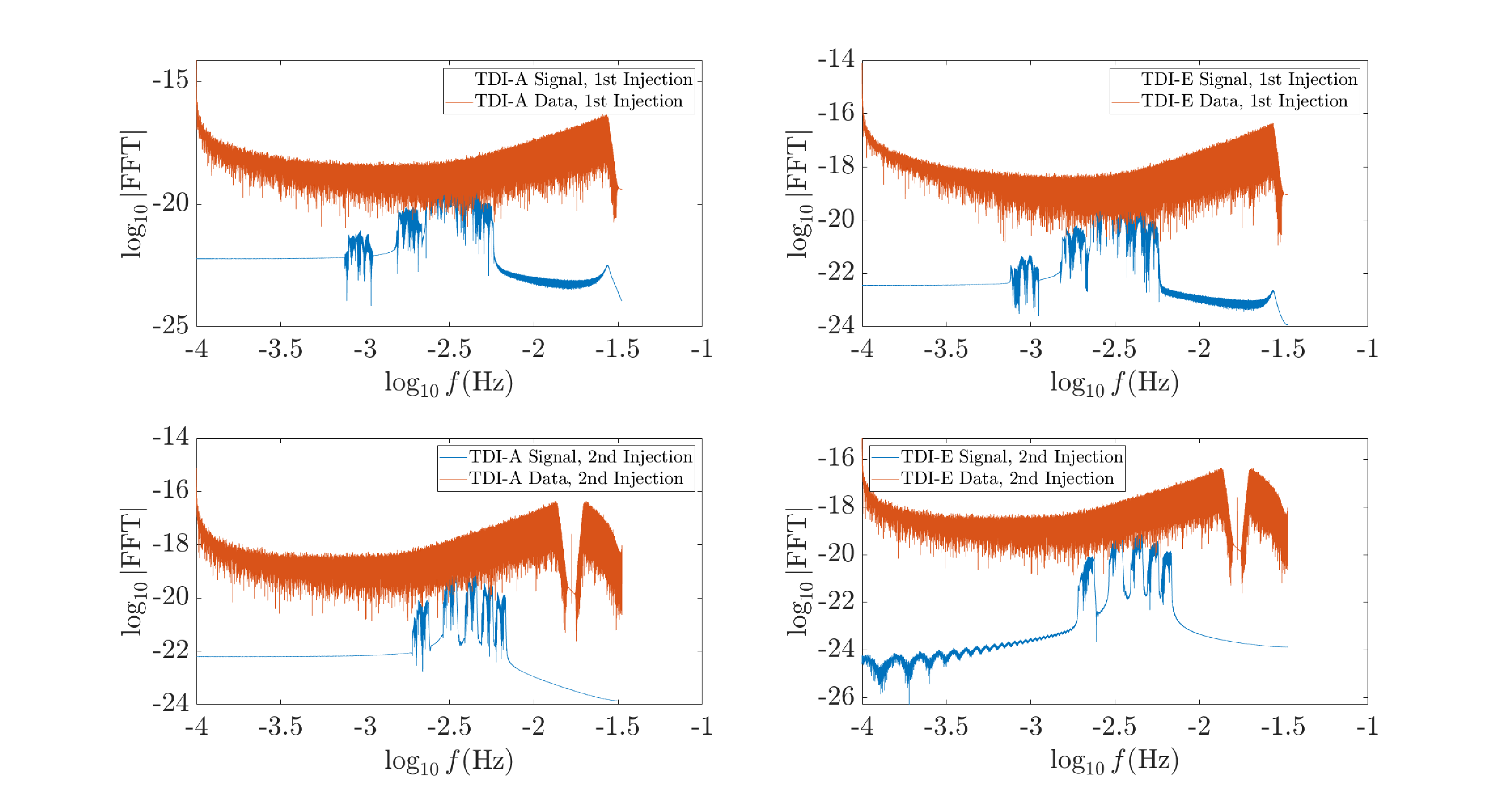}
	\caption{Spectrum of the two injected signals and simulated datasets.}
	\label{Fig-spectrum-injection}  
\end{figure}

In this paper, we inject two signals into stationary Gaussian noise generated by \texttt{LISACode}~\cite{Petiteau:2008zz}, with a duration of $0.5$-year, a cadence of $15$ seconds, and an SNR of $50$\textemdash to test the validity of the hierarchical search strategies. Their spectrum of TDI channels $A$ and $E$ for the noisy data and the noiseless signal are shown in Fig.~\ref{Fig-spectrum-injection}. It can be seen that the signals are at least one order of magnitude weaker than the noise, necessitating matched filtering to extract such weak signals by accumulating sufficient SNR.

For a real hierarchical search, the initial prior range should be as large as possible to encompass all potential EMRI sources predicted by the population model. Furthermore, based on the range used to generate the LDC‑$1.2$ injection signal, the initial wide search ranges for the phase-coupled parameters and the sky location are listed in Tab.~\ref{Search_range}.

\begin{table}[htb]
	\centering
	\begin{tabular}{l|c|c}
		\hline
		Parameter   &$1$st Injection   &$2$nd Injection \\
		\hline
		$\mu(M_\odot)$   &$[10,30]$     &$[3,30]$  \\
		$M(10^6M_\odot)$ &$[0.9,3]$     &$[0.5,5]$  \\
		$\lambda$(rad)   &$[0,\pi]$     &$[0,\pi]$  \\
		$S/M^2$          &$[0.5,0.9999]$   &$[0.5,1]$  \\
		$e_0$  &$[0.2,0.3]$    &$[0.1,0.4]$  \\
		\hline
		$\nu_0$(lags)    &$[-N,N]$    &$[-N,N]$  \\
		\hline
		$\theta_s$(rad)    &$[0,\pi]$     &$[0,\pi]$  \\
		$\phi_s$(rad)      &$[0,2\pi]$    &$[0,2\pi]$ \\	
		\hline
		%SNR   &$50$ &$50$ \\
		%\hline
	\end{tabular}
	\caption{The initial search range for the two injections, $N=0.5$-year$/15$ sec$=1048576$.}\label{Search_range}
\end{table}
% fiducial nu0
In this paper, we set the fiducial $\nu_0$ value to be $10^4$ lags ahead of its true value for both injections in the $7$D likelihood, corresponding to approximately $2$ days at a cadence of $15$ seconds. This implies that the detector captures the signal about $2$ days after it begins operating.  To obtain this fiducial $\nu_0$, we integrate the orbital ODEs backward for $10^4$ iterations starting from $t_0$ associated with the true value of $\nu_0$. Since the inspiral evolves slowly over timescales of a few days, choosing the fiducial $\nu_0$ that is $10^4$ lags earlier than the true value is not sufficiently general for arbitrary EMRI sources. A more detailed discussion on how to select an appropriate fiducial $\nu_0$ to retain as much SNR as possible is warranted, but this lies beyond the scope of the present work and is deferred to future studies.
A notable difference between the two injections is that the $1$st lies near the boundary of the search space, whereas the $2$nd lies near its center. The exact injection values are provided below for direct comparison with the best‑fit recovered values.

\subsection{$7$D versus $8$D} 
We have constructed the $7$D and $8$D likelihoods in the \hyperref[Reduced-Dim-LLR]{reduced-dimensionality-likelihood discussion} in the main text. The signal morphology exhibits a sharp $6$D  peak for the $8$D likelihood and a sharp $5$D peak for the $7$D likelihood. Owing to its lower dimensionality, the $7$D likelihood facilitates the identification of the peak pattern described in the \hyperref[Hierarchical-Search-Strategy]{hierarchical-search strategy}. Consequently, the $7$D likelihood is qualitatively better suited for the hierarchical search.
We also conduct a comprehensive quantitative comparison between the $7$D and $8$D likelihoods for the performance of searching the two injection signals illustrated in Fig.~\ref{Fig-spectrum-injection}.

\begin{figure*}[htbp]
	\centering 
	\includegraphics[width=\textwidth,height=3.7in,keepaspectratio]{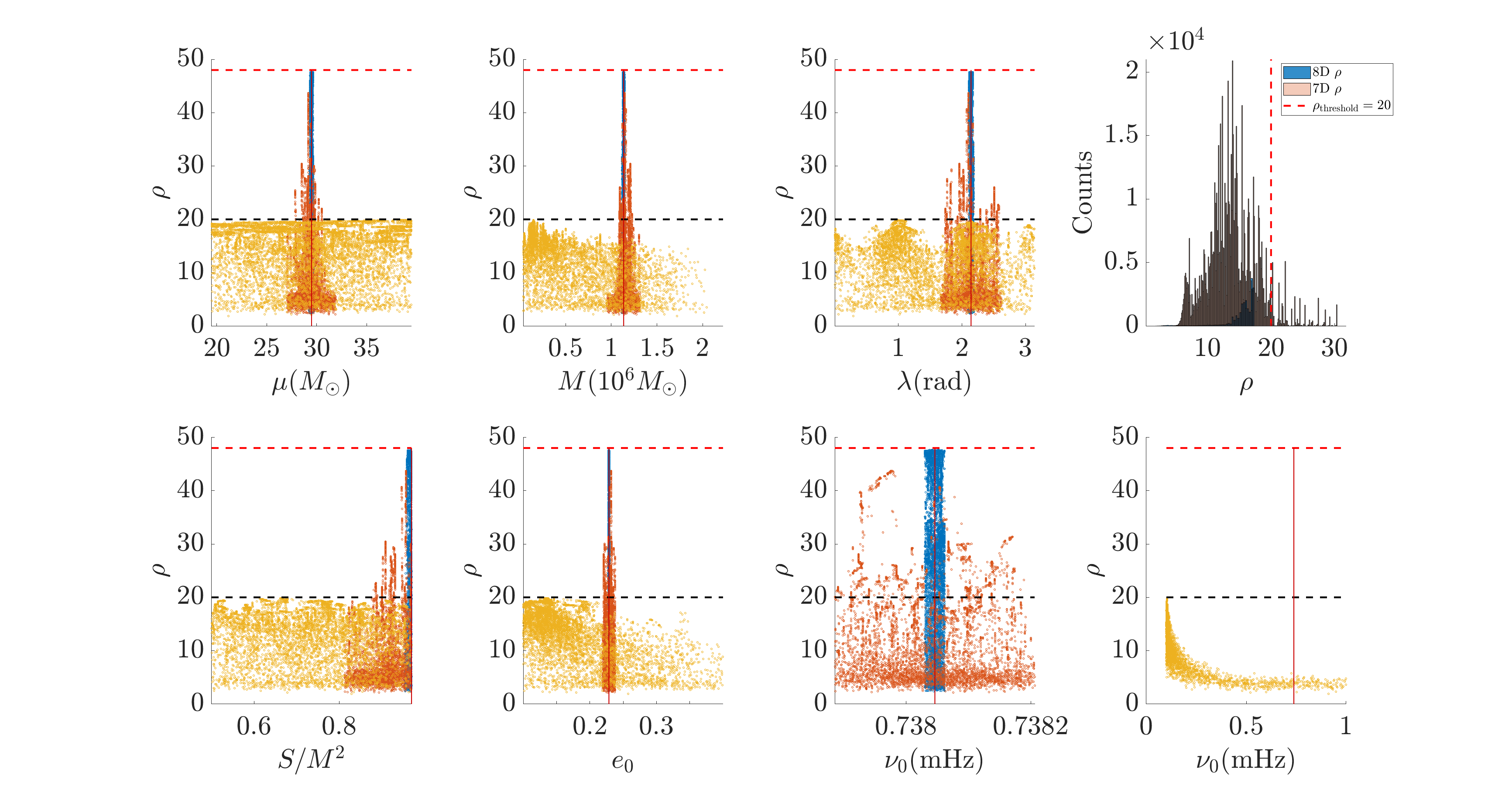}
	\caption{Comparison of the $7$D and $8$D likelihoods for the injection $1$. The black and red dashed horizontal lines correspond to $\rho_{\rm threshold}$ and $\rho_{\rm true}$, respectively. The true location is indicated by the red vertical line.}
	\label{Fig:7D_vs_8D_p1}  
\end{figure*}

The results of the injection $1$ are summarized in Fig.~\ref{Fig:7D_vs_8D_p1}. Columns 1\textemdash 3 show the $1$D slices of the collected peaks obtained with the $8$D likelihood under three different search ranges. These ranges are defined using the Fisher information matrix (FIM) uncertainty $\overline{\sigma}$
at the injection $\overline{x}_0$: a narrow range $\overline{x}_0\pm 5\overline{\sigma}$, a moderate range $\overline{x}_0\pm 50\overline{\sigma}$, and the full wide range listed in Tab.~\ref{Search_range}, corresponding to blue, orange and yellow scatter points, respectively. These results confirm the expectation that a direct search works well for small search ranges but gradually fails as the range increases, with the landmark peak losing SNR.  For the $1$D peak slice of $\nu_0$, the most sensitive phase-coupled parameter, we provide a zoom-in view in Row $2$, Column $3$ and the original view in Row $2$, Column $4$. Row $1$, Column $4$ compares the peak LLR $\rho$ (defined in Eq.~(\ref{eq:8D-LLR})) between the $7$D and $8$D likelihoods using the wide search range from Tab.~\ref{Search_range}. It can be seen that the $7$D likelihood is able to extract secondary peaks with $\rho>20$, whereas the $8$D likelihood cannot. If we set a GW-induced threshold of $\rho_{\rm threshold}=20$ following~\cite{Chua:2021aah}, this indicates that the $8$D likelihood loses all GW-induced peaks and therefore cannot be applied in the hierarchical search.

% boundary case

\begin{figure*}[htbp]
	\centering 
	\includegraphics[width=\textwidth,height=3.7in,keepaspectratio]{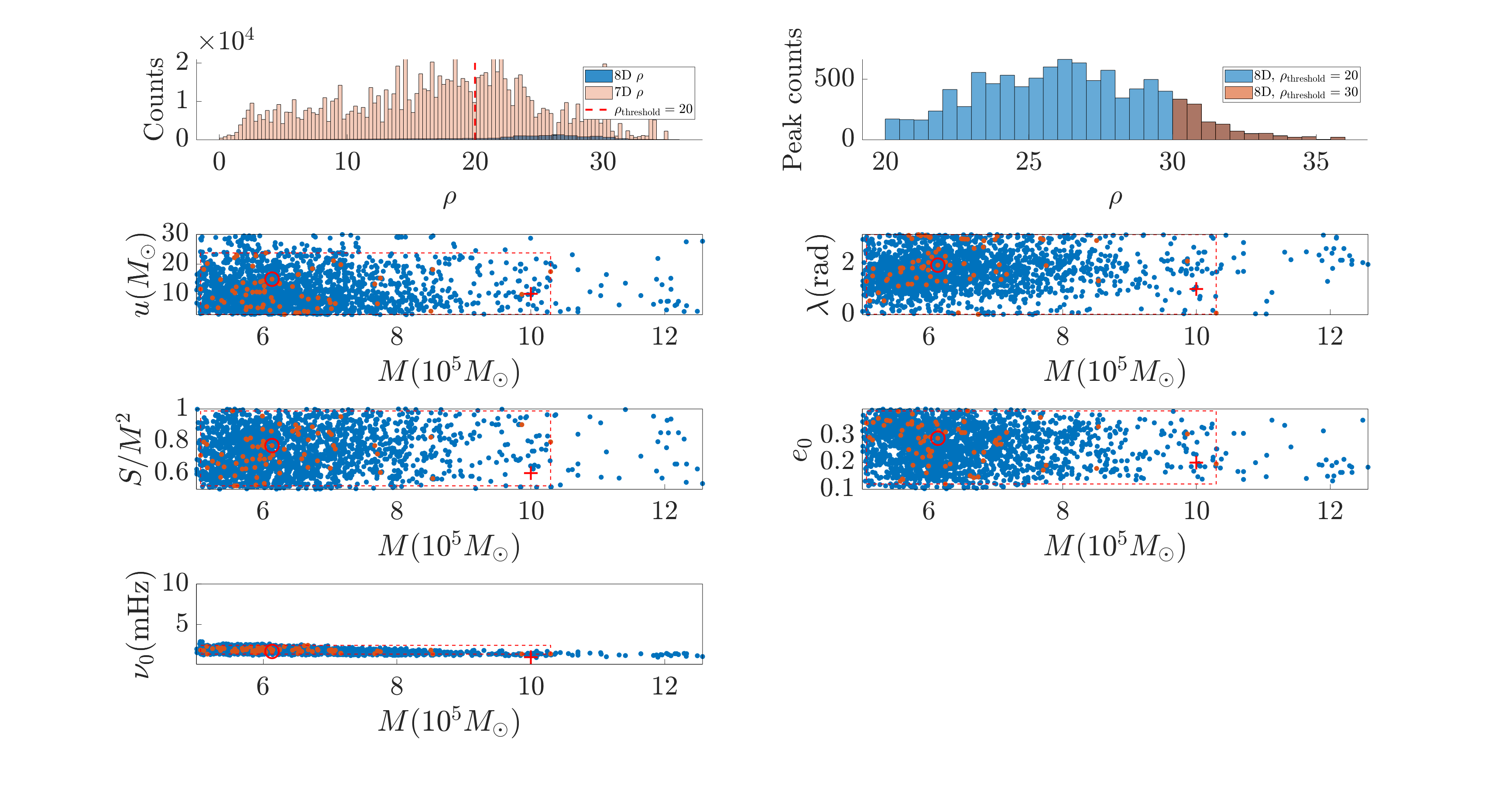}
	\caption{Comparison of the $7$D and $8$D likelihoods for the injection $2$. The blue and red scatter points denote the GW-induced and high-LLR peaks, respectively. The red circle marks the landmark peak with the highest LLR, and the red plus sign indicates the injection. The dashed red square represents the smaller region estimated from the high-LLR peaks.}
	\label{Fig:7D_vs_8D_p2}  
\end{figure*}

The results of the injection $2$ are presented in Fig.~\ref{Fig:7D_vs_8D_p2}. Row $1$ shows the histogram of peak $\rho$. Specifically, Row $1$, Column $1$ demonstrates that both the $8$D and $7$D likelihoods are capable of identifying GW-induced peaks, while Row $1$, Column $2$ displays the GW-induced peaks alongside those with higher LLR ($\rho>30$). The remaining subfigures, which illustrate the $2$D peak slices, indicate that the peaks obtained with the $8$D likelihood are significantly biased. This bias arises because the signal morphology corresponds to a $6$D sharp peak, and the presence of the most sensitive parameter $\nu_0$ obscures the peak pattern discussed in the \hyperref[Hierarchical-Search-Strategy]{hierarchical-search strategy}.

% central case

In summary, we conclude that the $7$D likelihood can consistently reveal the peak pattern discussed in the \hyperref[Hierarchical-Search-Strategy]{hierarchical-search strategy} for both injections. This is because it estimates $\nu_0$ via a lag shift, thereby reducing the signal morphology from $6$D to $5$D, whose simpler characteristics make it more conducive to revealing the peak pattern.
% -------------------------------------
\subsection{Tuning the $\rho_{\rm threshold}$  for a real hierarchical search} 
\label{rho_threshold}
During each iteration, we select high-LLR peaks from the GW-induced peaks by applying a threshold $\rho_{\rm threshold}$, retaining only those with $\rho>\rho_{\rm threshold}$. To tune the hierarchical search, the threshold is chosen such that a sufficient number of secondary peaks are retained to render the peak pattern clearly discernible. In general, the number of secondary peaks per iteration should be as large as computationally feasible. However, as described in Section~\ref{app:cost}, the independent PSO searches are executed serially in our CPU pipeline, and each search takes nearly a day. Consequently, we cannot afford too many independent PSO searches per iteration in the hierarchical search. The values of  $\rho_{\rm threshold}$ and the corresponding numbers of high-LLR peaks are listed in Tab.~\ref{tab:rho-threshold}.
\begin{table}[htb]
	\centering
	\begin{tabular}{ccccccccc}
		\hline\hline
		Iteration index &$1$ &$2$ &$3$ &$4$   &$5$ &$6$ &$7$ &$8$ \\
		\hline
		Injection $1$ &\makecell{$30$\\$3212$} &\makecell{$30$\\$2809$} &\makecell{$32$\\$10194$} &\makecell{$38.5$\\$2700$}   &\makecell{$37$\\$10751$} &\makecell{$46.4$\\$2008$} &\makecell{$46.8$\\$1766$} &$\sim$\\
		\hline
		Injection $2$ &\makecell{$20$\\$4860$} &\makecell{$31$\\$1030$} &\makecell{$38.5$\\$696$} &\makecell{$40$\\$2502$}   &\makecell{$40$\\$4858$} &$\sim$ &$\sim$ &$\sim$ \\
		\hline\hline
	\end{tabular}
	\caption{Choice of $\rho_{\rm threshold}$(top) and the corresponding number of high-LLR peaks(bottom) used to estimate the reduced search region for the next iteration.}\label{tab:rho-threshold}
\end{table}
To mitigate the insufficient number of high-LLR peaks, we employ the supplementary strategy introduced in Fig.~\ref{Fig:tricks-hierarchical-search} and in the \hyperref[Hierarchical-Search-Strategy]{hierarchical-search discussion}.

The fact is that a larger number of high-LLR peaks yields a clearer peak pattern, thereby facilitating the hierarchical search. Consequently, for practical hierarchical searches, GPU should be introduced to substantially improve pipeline efficiency, making it easier to obtain a sufficient number of secondary peaks.

%--------------------------------------
\subsection{Choosing the proper fiducial $\nu_0$}
% 2026/08/05 ZXB
For illustration, we set the fiducial $\nu_0$ to $10,000$ lags before its true value for both injections in the 7D likelihood, meaning the detector captures the signal about $2$ days after operation start (at a $15$‑second cadence).
Note that the $7$D likelihood reveals the peak pattern more stably and clearly, since the peak morphology simplifies as its dimensionality decreases to $5$D. Therefore, the $7$D likelihood is necessary to initiate the hierarchical search, in which the choice of the fiducial  $\nu_0$ presents a fine-tuning problem. Since the inspiral evolves slowly, the variation of  $\nu_0$ over a timescale of weeks or months is small. If the fiducial  $\nu_0$ is set too early relative to its true value, the initial segment of the signal will be missing. Consequently, the closer the fiducial  $\nu_0$ is to the true value, the larger the fraction of the signal that can be retained. This poses a challenge for a fully blind search. Further analysis of the EMRI population model and the sensitivity band of the detector is required for an optimal choice of the fiducial $\nu_0$.
% -------------------------------------
\subsection{The peak pattern} 
\label{peak-pattern}
\begin{figure*}[htbp]
		\centering 
		\includegraphics[width=\textwidth,height=3.7in,keepaspectratio]{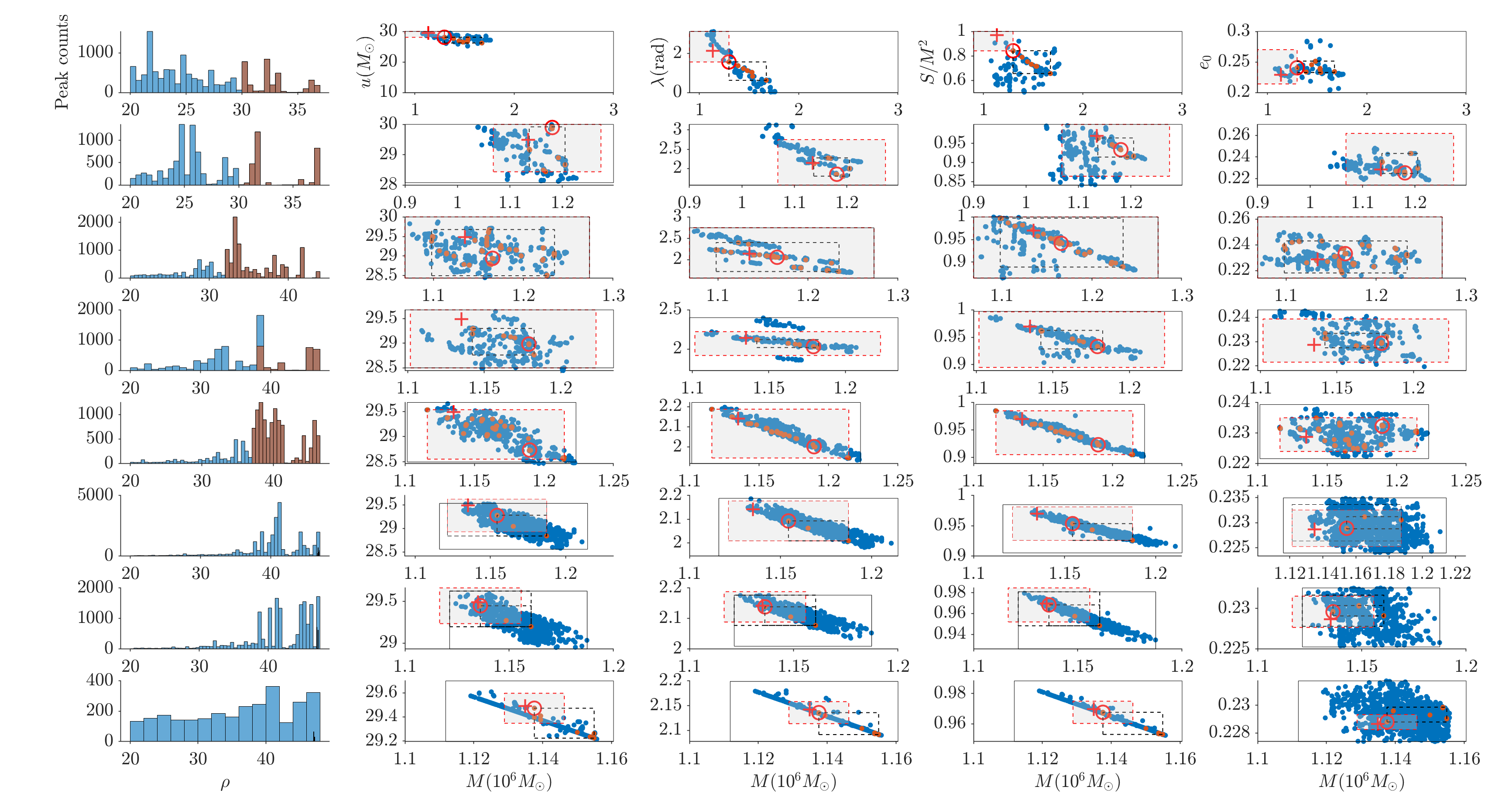}
		\caption{Peak pattern for the injection $1$. Blue and red scattered points denote GW-induced and high-LLR peaks, respectively. The red circle marks the landmark peak (highest LLR), and the red plus sign marks the injection. Dashed black squares show the smaller range directly estimated from high-LLR peaks, and the shaded regions enclosed by the dashed red squares are extended and shifted region (see Fig.~\ref{Fig:tricks-hierarchical-search} and the \hyperref[Hierarchical-Search-Strategy]{hierarchical-search strategy}). The black square is the search region input for this iteration's hierarchical search.}
		\label{Fig-hie-p1}  
	\end{figure*}

\begin{figure*}[htbp]
		\centering 
		\includegraphics[width=\textwidth,height=3.7in,keepaspectratio]{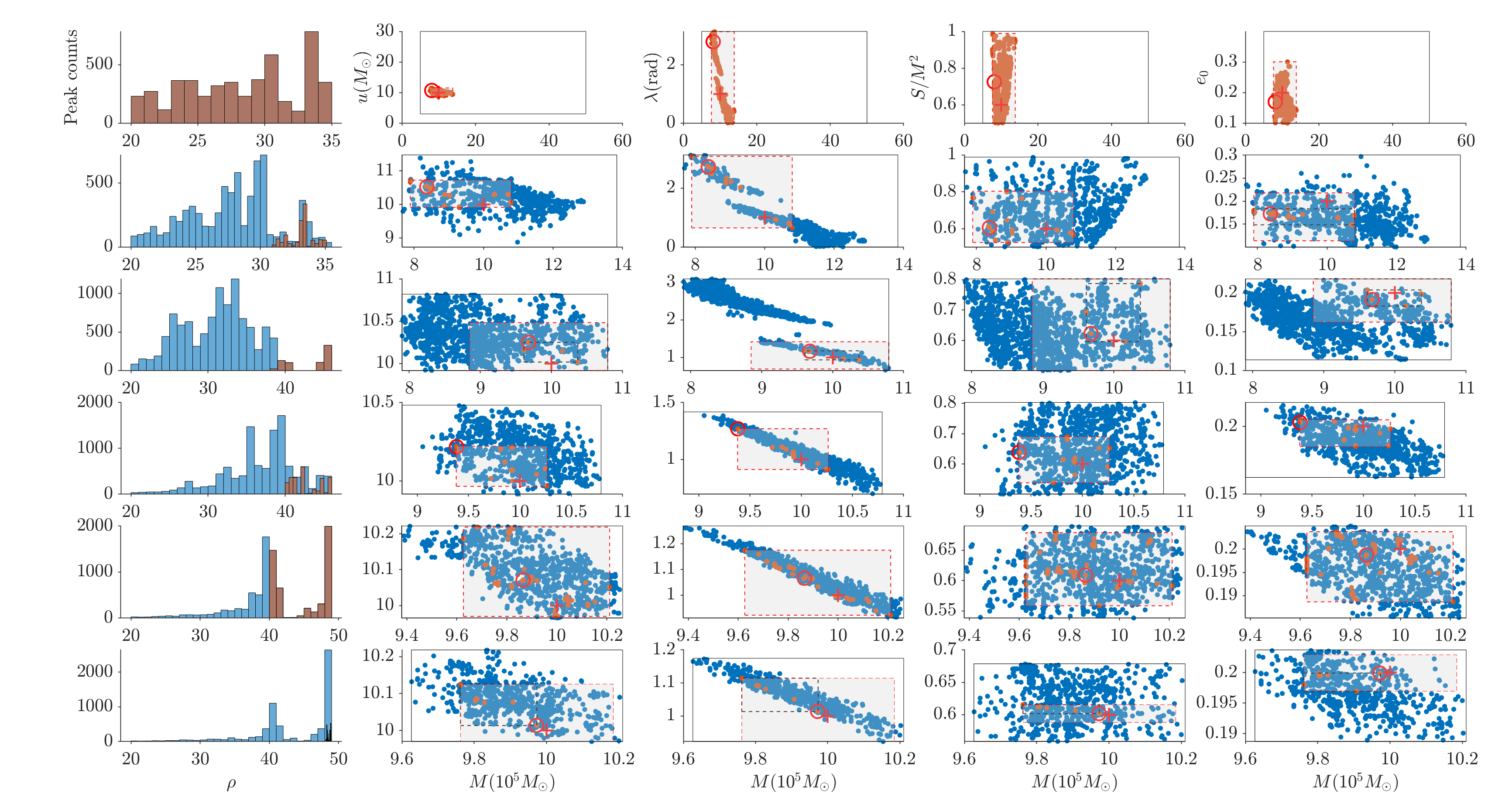}
		\caption{Peak pattern of the injection $2$ with the same conventions as Fig.~\ref{Fig-hie-p1}.}
		\label{Fig-hie-p2}  
	\end{figure*}

More detailed illustrations of the hierarchical searches for the two injections are provided in Fig.~\ref{Fig-hie-p1} and Fig.~\ref{Fig-hie-p2}, respectively. Each row corresponds to one iteration. The first column displays the peak histogram, showing the distinction between low-LLR and high-LLR peaks, while the remaining columns depict the estimation of the reduced search ranges based on the clustering behavior of high-LLR peaks, following the peak pattern discussed in the \hyperref[Hierarchical-Search-Strategy]{hierarchical-search strategy}.

% injection 1, round 8, linear fitting; 
For the injection $1$, as the hierarchical search progresses, the linear peak pattern for the pairs $(\mu,M),(\lambda,M)$, and $(S/M^2,M)$ becomes increasingly clear. Therefore, to accelerate the hierarchical search, the PSO searches in the $8$th iteration are conducted along the lines fitted from the high-LLR peaks in the previous iteration (see Row $8$ of Fig.~\ref{Fig-hie-p1}).

%  injection 2, round 4, no neighbours 
For the injection $2$, we introduce the neighbors of the identified peak region for $e_0$ only in the $2$nd iteration (Row $2$ of Fig.~\ref{Fig-hie-p2}), and directly use the identified peak region in the $4$th iteration (Row $4$ of Fig.~\ref{Fig-hie-p2}) to estimate the reduced search range.  The former approach accounts for the fact that  $e_0$ is an initial parameter for orbital evolution and is therefore too sensitive to be localized precisely.  In the latter case, the landmark peak is located at the corner of the identified peak region, corresponding to case (b) illustrated in Fig.~\ref{Fig:tricks-hierarchical-search}. In general, the neighbors of the identified peak region should be used to ensure that the primary peak is enclosed, as discussed in the \hyperref[Hierarchical-Search-Strategy]{hierarchical-search strategy}. However, to save computational cost, we directly use the identified peak region\textemdash which indeed encloses the known injection\textemdash as the estimated reduced search range at the $4$th iteration, as a preliminary verification of the hierarchical search strategy. Note that this simplification somewhat compromises the generality of the hierarchical search for the $2$nd injection. Nevertheless, the hierarchical search strategy based on the peak pattern proves to be general, so that additional iterations beyond the current $6$ would still bring the landmark peak progressively closer to the primary peak.

% (u,M) 1st iteration, 
As shown in Row $1$, Column $2$ of Fig.~\ref{Fig-hie-p1} and Fig.~\ref{Fig-hie-p2}, the search ranges of the two masses $(\mu,M)$ are significantly reduced after the $1$st iteration of the hierarchical search for both injections. This demonstrates that the hierarchical search can effectively handle arbitrarily large initial search ranges for this pair of parameters across all possible EMRI sources.

In summary, the hierarchical search strategy proves effective at each iteration for both injections, confirming the generality of the peak pattern. This demonstrates that our hierarchical search strategy has the potential to deliver accurate parameter estimation across all EMRI sources.

	% hierarchical injection, detailed look
	% -------------------------------------
\section{Illustration of the local degeneracy} 
\label{local-degeneracy}
\begin{figure*}[htbp]
		\centering 
		\includegraphics[width=\textwidth,height=3.7in,keepaspectratio]{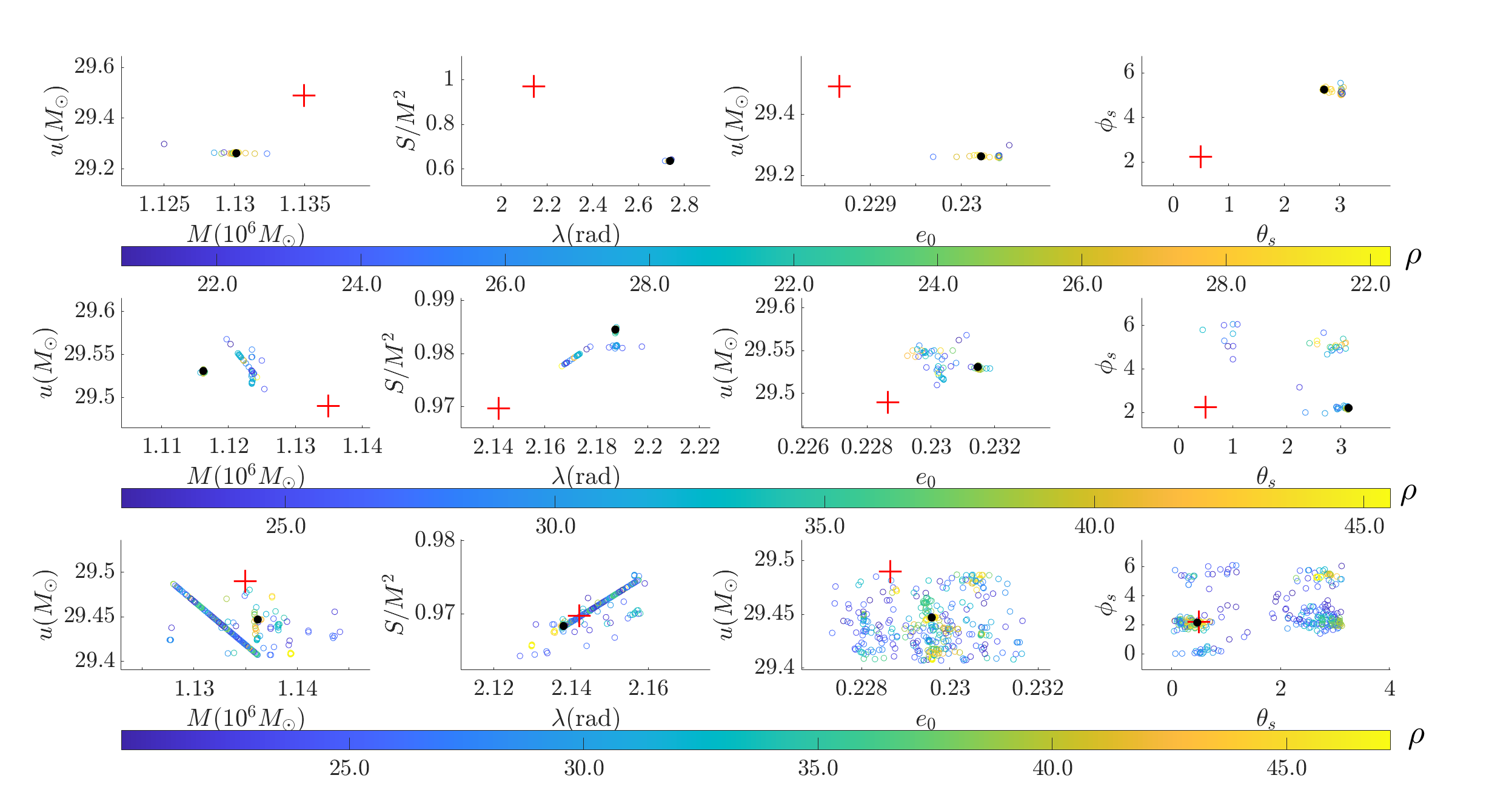}
		\caption{Local peak pattern for the injection $1$. The representative peak is denoted by black filled points, and the red plus sign marks the injection. The $\rho$ values in the color bar are defined in Eq.~(\ref{eq:8D-LLR}).}
		\label{Fig-local-p1-1}  
	\end{figure*}

\begin{figure*}[htbp]
		\centering                                
		\includegraphics[width=\textwidth,height=3.7in,keepaspectratio]{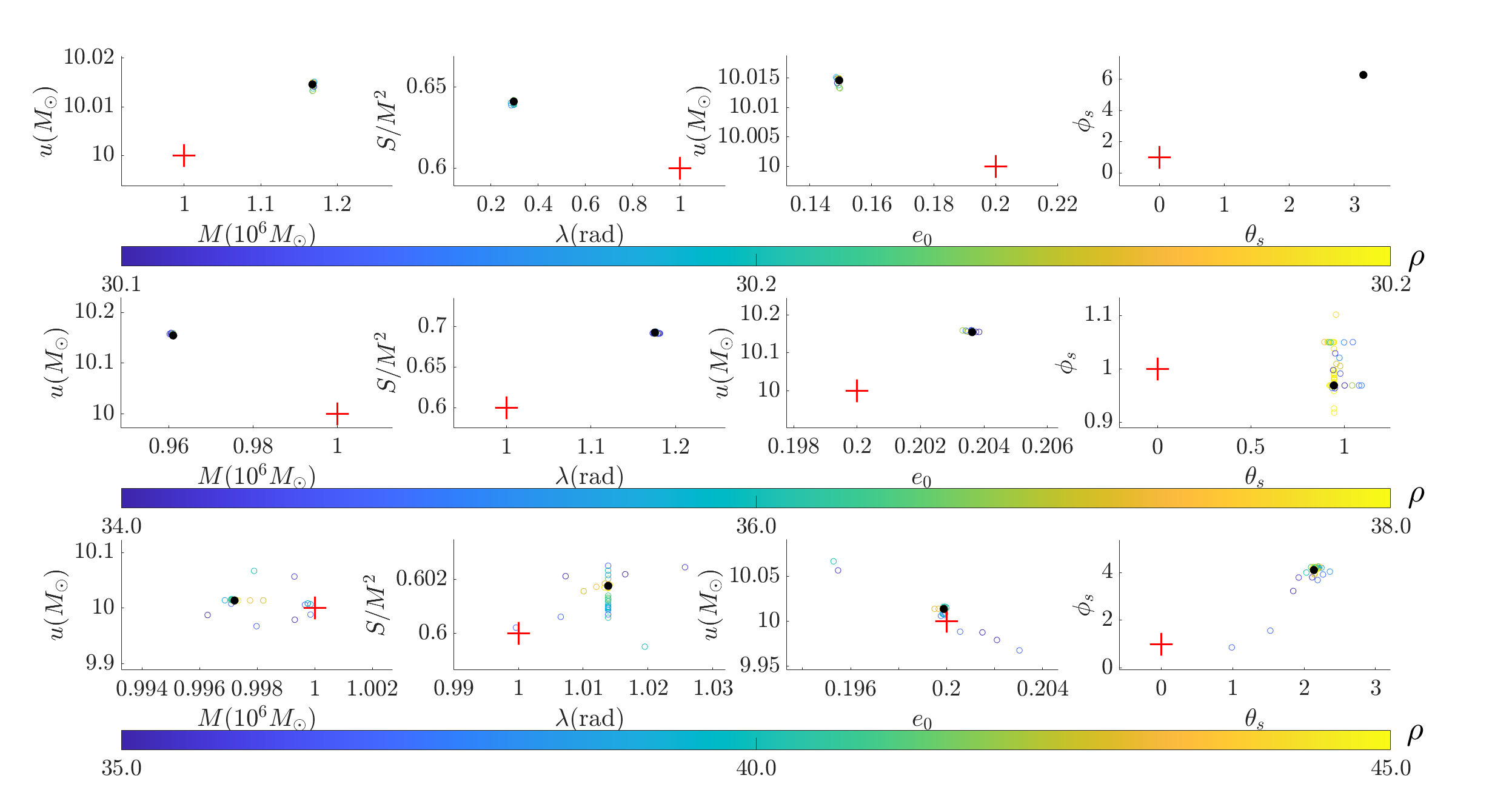}
		\caption{Local pattern of the peaks for injection $2$ with the same settings as Fig.~\ref{Fig-local-p1-1}.}
		\label{Fig-local-p2-1}  
	\end{figure*}

% local cluster, secondary peak
The local peak patterns around three representative secondary peaks are visualized. These peaks correspond to local maxima at $\rho=29.14,~39.84,~47.22$ for the $1$st injection and $\rho=30.34,~39.56,~48.89$ for the $2$nd injection. Fig.~\ref{Fig-local-p1-1} and Fig.~\ref{Fig-local-p2-1} show the 2D slices of all peak scatter points within the reduced range $\overline{x}_0\pm 25 \overline{\sigma}$ for the two injections, respectively, where $\overline{x}_0$ denotes the location of the representative secondary peak and $\overline{\sigma}$  is its local FIM uncertainty estimate and rows $1-3$ correspond to increasing LLR of the representative secondary peaks. Three observations emerge: First, as the LLR increases, the local cluster moves closer to the injection. Second, within the reduced search range, a peaked pattern persists, with the representative secondary peak serving as the local maximum. Third, the degenerate sky-location becomes increasingly distinct as the representative peak approaches the primary peak.

\begin{figure*}[htbp]
		\centering 
		\includegraphics[width=\textwidth,height=3.7in,keepaspectratio]{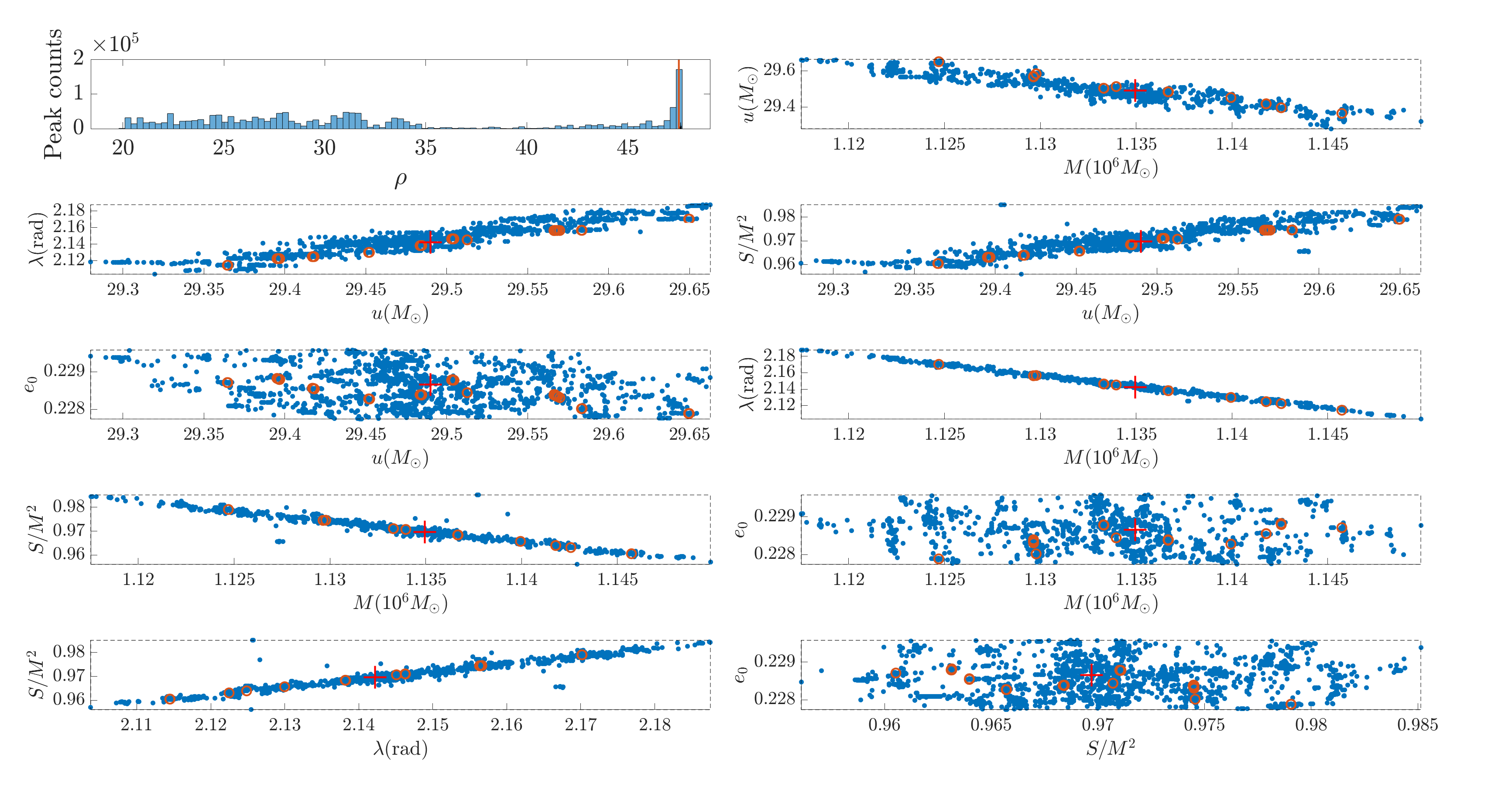}
		\caption{Local degenerate forest for the injection $1$. Blue scatter points denote the GW-induced peaks; orange scatter points represent degenerate peaks with  $\rho>\rho_{\rm true}$. The value of $\rho_{\rm true}$ is shown in the peak histogram in Row $1$, Column $1$.}
		\label{Fig-local-p1-2}  
	\end{figure*}

\begin{figure*}[htbp]
		\centering                                
		\includegraphics[width=\textwidth,height=3.7in,keepaspectratio]{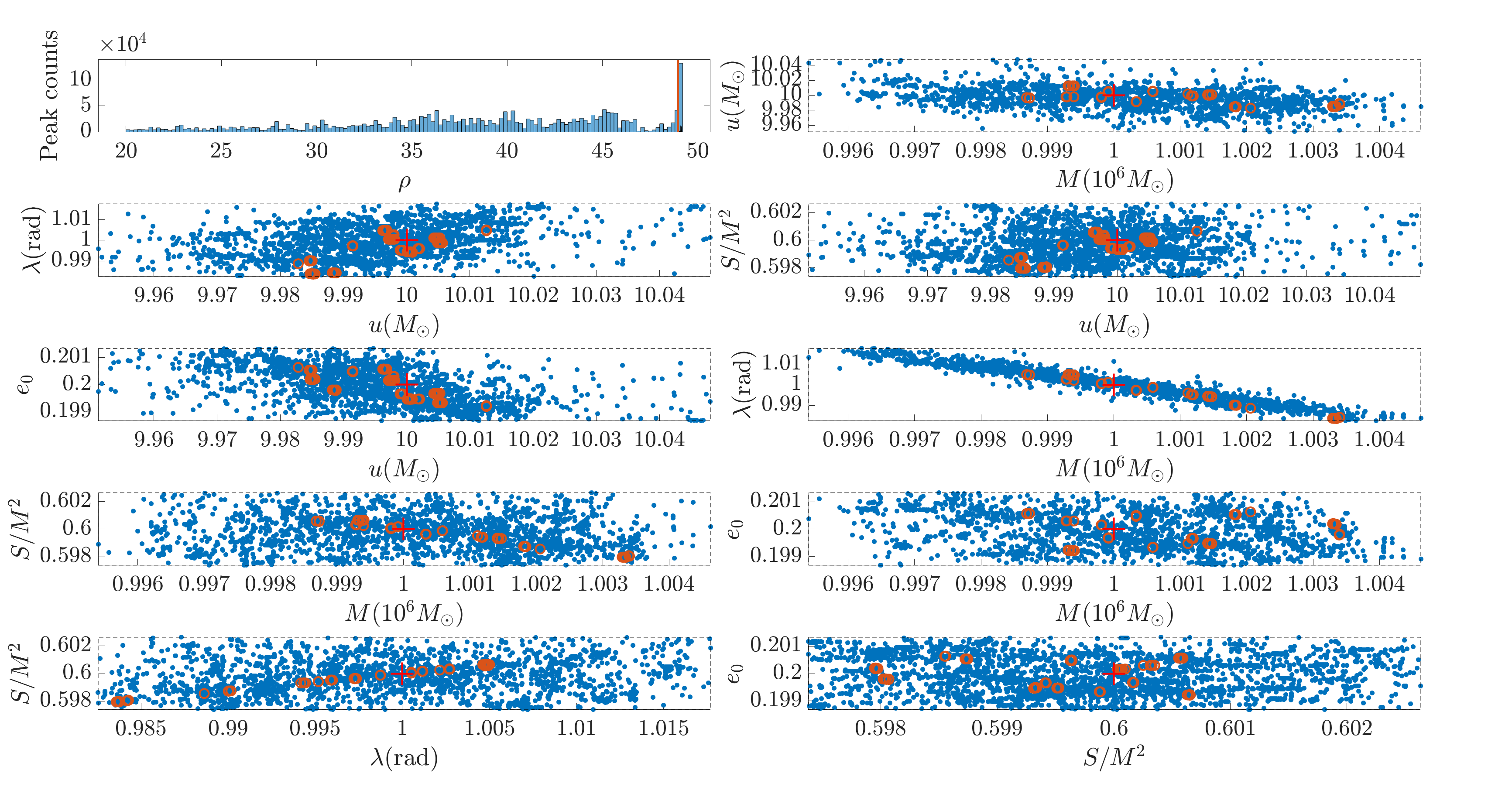}
		\caption{Local degenerate forest for the injection $2$ with the same settings as Fig.~\ref{Fig-local-p1-2}.}
		\label{Fig-local-p2-2}  
	\end{figure*}

% local forest, primary peak
The local pattern around the primary peak (i.e., the injection) is illustrated in Fig.~\ref{Fig-local-p1-2} and Fig.~\ref{Fig-local-p2-2} for the two injections, respectively, using a small search range of $\overline{x}_{\rm true}\pm 10 \overline{\sigma}$ where $\overline{\sigma}$ is the FIM uncertainty at the injection. A dense local degenerate forest emerges, containing tens of thousands of distinctly clustered peaks with $\rho>\rho_{\rm true}$. In some sub-dimensions, the peaks exhibit a linear spatial distribution\textemdash for example, in the $2$D slices of $(\lambda,M)$ for the GW-induced peaks, and in the 2D slices of $(\lambda,M,S/M^2)$ for the degenerate forest. Peaks in $e_0$-related sub-dimensions fluctuate constantly, as $e_0$ is an initial parameter governing orbital evolution; consequently, its sensitivity precludes precise localization. For the other phase-coupled parameters, the linear pattern depends on either the EMRI source parameters or the location of the injection in the search space. 

% consistent with the concepture 
In summary, we confirm the peaked structure within each cluster and the multi-peaked landscape featuring a dominant central peak, as qualitatively illustrated in Fig.~\ref{fig:peak_landscape_1} and discussed in the \hyperref[Hierarchical-Search-Strategy]{hierarchical-search strategy}. The pattern becomes more evident as more peaks are collected. Regarding the origin, the multi-peaked landscape arises from different deviations in the phase-coupled parameters from their true values, leading to varying degrees of phase match, while the peaked structure within each cluster results from deviations in the sky location for given deviations in the phase-coupled parameters.

	% local peak

	% local degenerate forest

	% -------------------------------------
\subsection{Illustration of nonlocal degeneracy}
\begin{figure*}[htbp]
		\centering 
		\includegraphics[width=\textwidth,height=3.7in,keepaspectratio]{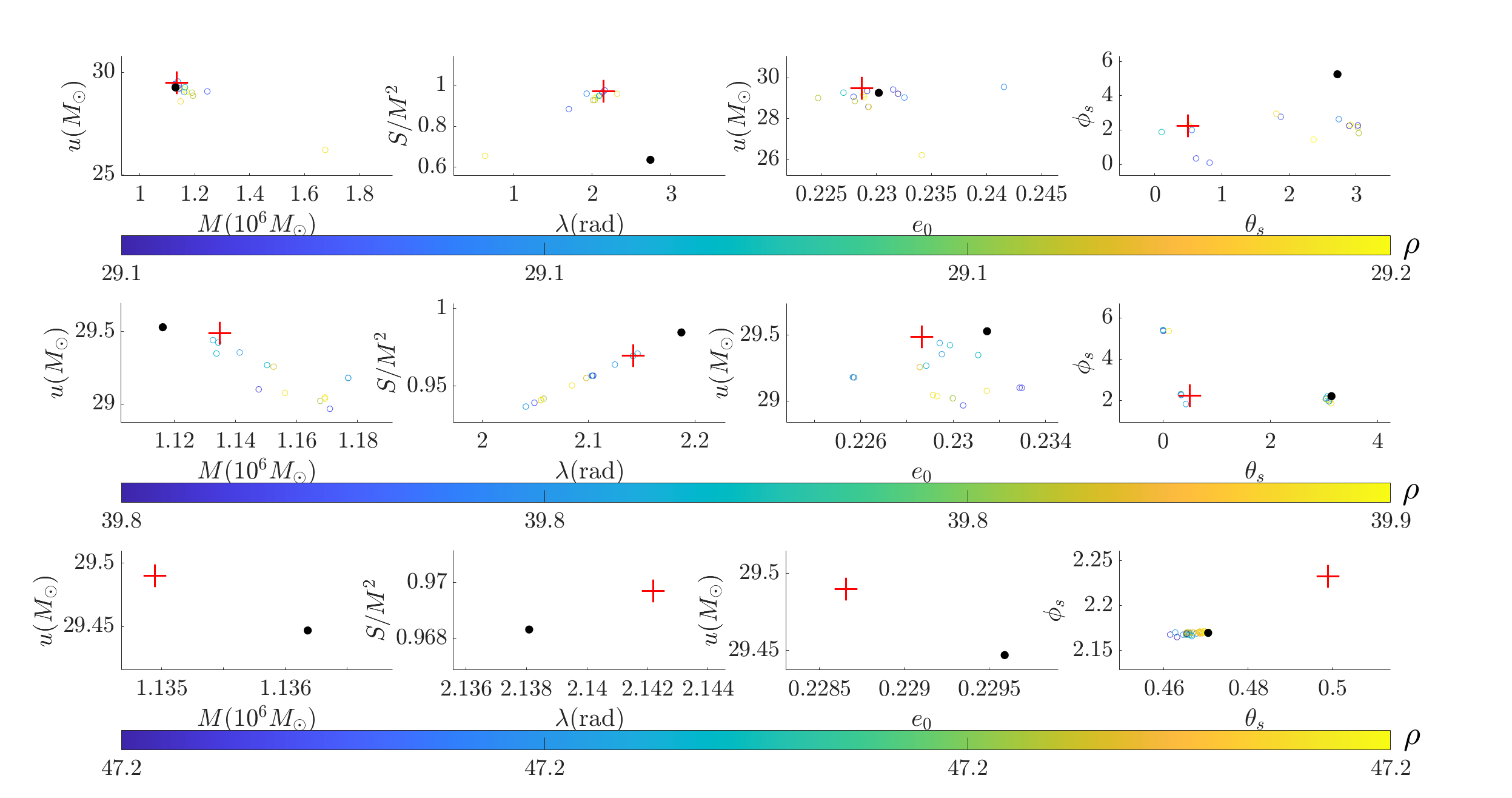}
		\caption{Nonlocal degeneracy pattern of the peaks for the $1$st injection, with the same settings as in Fig.~\ref{Fig-local-p1-1}. }
		\label{Fig-no-local-p1}  
	\end{figure*}

\begin{figure*}[htbp]
		\centering                                
		\includegraphics[width=\textwidth,height=3.7in,keepaspectratio]{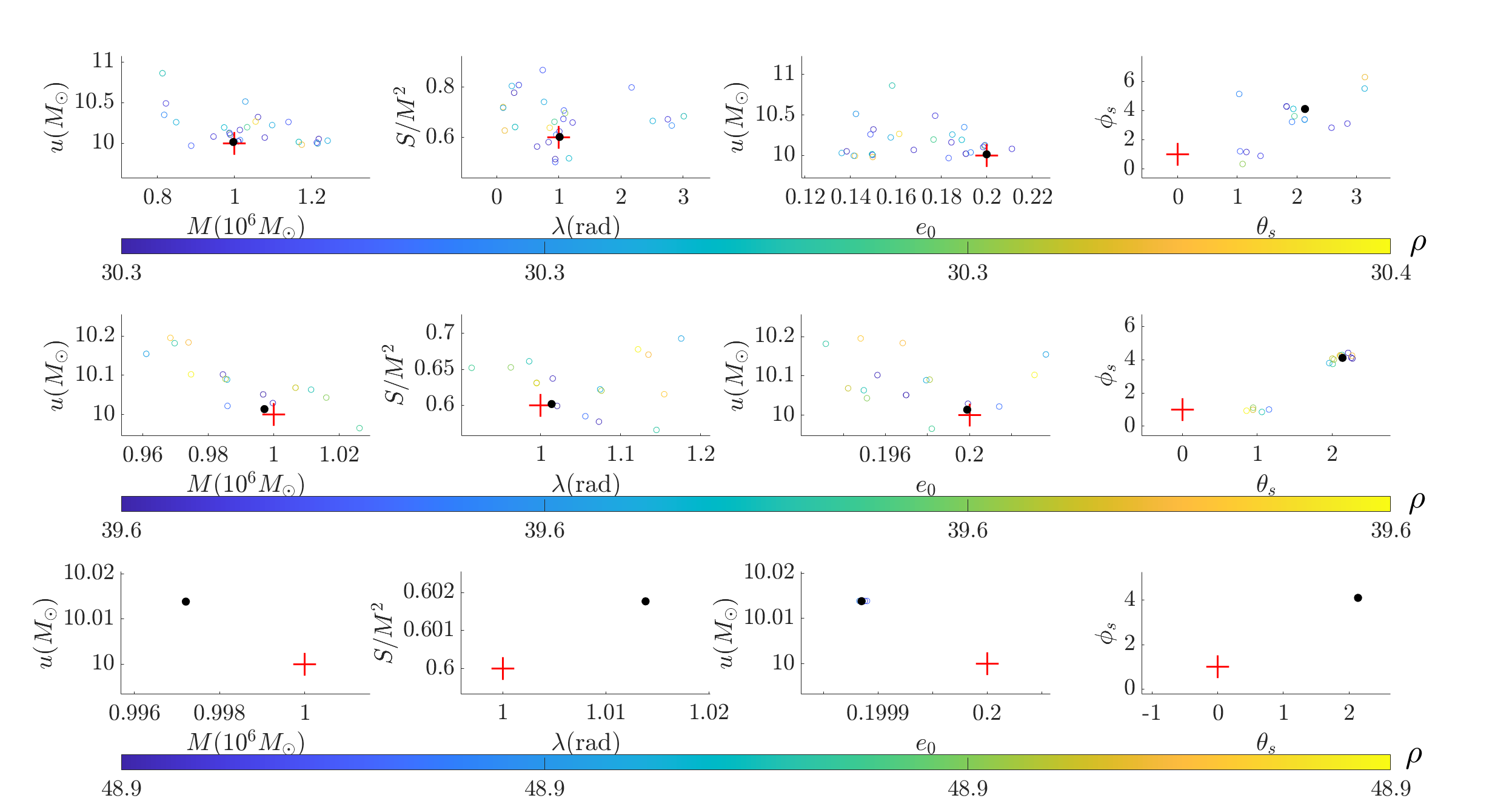}
		\caption{Nonlocal degeneracy pattern of the peaks for the $2$nd injection with the same settings as Fig.~\ref{Fig-local-p1-1}. The degenerate sky-location is observed.}
		\label{Fig-no-local-p2}  
	\end{figure*}

The nonlocal degeneracy around the same three representative peaks as in Sec.~\ref{local-degeneracy} is visualized. For each representative peak, we select all peaks whose LLR differs from that peak by less than $0.01$ over the entire wide search range. The results are illustrated in Fig.~\ref{Fig-no-local-p1} and Fig.~\ref{Fig-no-local-p2} for the two injections, respectively. Strong nonlocal degeneracy is observed in the EMRI likelihood surface, arising from the alignment of dominant and subdominant harmonics, as discussed in~\cite{Chua:2021aah}. This degeneracy becomes weaker as the representative peak approaches the primary one.
	% no-local degenerate

\clearpage
\onecolumngrid
\section{State of the Art in EMRI Searches}
\label{app:prior-work}
\begin{table}[H]
\centering
\scriptsize
\setlength{\tabcolsep}{2.2pt}
\renewcommand{\arraystretch}{1.12}
\begin{tabular}{@{}lllllll@{}}
\hline\hline
\parbox[t]{0.10\textwidth}{\raggedright Study\par} & \parbox[t]{0.13\textwidth}{\raggedright Demonstrated source model\par} & \parbox[t]{0.08\textwidth}{\raggedright Physical dimension\par} & \parbox[t]{0.16\textwidth}{\raggedright Search statistic\par} & \parbox[t]{0.17\textwidth}{\raggedright Initial prior scope\par} & \parbox[t]{0.09\textwidth}{\raggedright Sky treatment\par} & \parbox[t]{0.14\textwidth}{\raggedright Reported endpoint\par} \\
\hline
\parbox[t]{0.10\textwidth}{\raggedright Ye et al.~\cite{Ye:2023lok}\par} & \parbox[t]{0.13\textwidth}{\raggedright Schwarzschild eccentric EMRI\par} & \parbox[t]{0.08\textwidth}{\raggedright Reduced physical search\par} & \parbox[t]{0.16\textwidth}{\raggedright Physical-harmonic search followed by semicoherent phenomenological waveforms\par} & \parbox[t]{0.17\textwidth}{\raggedright Broad physical priors; no additional physical prior information\par} & \parbox[t]{0.09\textwidth}{\raggedright Sky angles included; weakly constrained\par} & \parbox[t]{0.14\textwidth}{\raggedright Signal identification and narrower physical-parameter ranges\par} \\

\parbox[t]{0.10\textwidth}{\raggedright Strub et al.~\cite{Strub:2025dfs}\par} & \parbox[t]{0.13\textwidth}{\raggedright Schwarzschild eccentric EMRI\par} & \parbox[t]{0.08\textwidth}{\raggedright 11 parameters\par} & \parbox[t]{0.16\textwidth}{\raggedright Time--frequency statistic followed by coherent SNR maximization\par} & \parbox[t]{0.17\textwidth}{\raggedright Wide mass and eccentricity priors; two-week time-to-plunge window\par} & \parbox[t]{0.09\textwidth}{\raggedright All sky in the follow-up stages\par} & \parbox[t]{0.14\textwidth}{\raggedright Maximum-likelihood recovery and MCMC posterior\par} \\

\parbox[t]{0.10\textwidth}{\raggedright Cole et al.~\cite{Cole:2025sqo}\par} & \parbox[t]{0.13\textwidth}{\raggedright Schwarzschild eccentric EMRI\par} & \parbox[t]{0.08\textwidth}{\raggedright 11 parameters\par} & \parbox[t]{0.16\textwidth}{\raggedright Sequential simulation-based inference (TMNRE)\par} & \parbox[t]{0.17\textwidth}{\raggedright Wide 11-dimensional prior\par} & \parbox[t]{0.09\textwidth}{\raggedright All-sky angles included\par} & \parbox[t]{0.14\textwidth}{\raggedright Prior-volume contraction and one-dimensional marginal proposals\par} \\

\parbox[t]{0.10\textwidth}{\raggedright Wang et al.~\cite{Wang:2026lcc}\par} & \parbox[t]{0.13\textwidth}{\raggedright Galactic-center XMRI; black-hole spin recovered\par} & \parbox[t]{0.08\textwidth}{\raggedright Source-targeted model\par} & \parbox[t]{0.16\textwidth}{\raggedright Semicoherent multi-harmonic $\mathcal{F}$-statistic with PSO\par} & \parbox[t]{0.17\textwidth}{\raggedright Galactic-center XMRI prior\par} & \parbox[t]{0.09\textwidth}{\raggedright Galactic center targeted\par} & \parbox[t]{0.14\textwidth}{\raggedright Targeted detection and parameter recovery\par} \\

\parbox[t]{0.10\textwidth}{\raggedright This work\par} & \parbox[t]{0.13\textwidth}{\raggedright Generic, spinning, eccentric, and inclined EMRI\par} & \parbox[t]{0.08\textwidth}{\raggedright Complete 14-parameter EMRI signal\par} & \parbox[t]{0.16\textwidth}{\raggedright Coherent profile likelihood: $7$D localization followed by $8$D refinement\par} & \parbox[t]{0.17\textwidth}{\raggedright Population-scale mass, spin, inclination, and eccentricity ranges\par} & \parbox[t]{0.09\textwidth}{\raggedright All sky from the first stage\par} & \parbox[t]{0.14\textwidth}{\raggedright End-to-end full-parameter maximum-likelihood recovery\par} \\
\hline\hline
\end{tabular}
\caption{Comparison with recent hierarchical EMRI search and inference demonstrations.}
\label{tab:prior-work}
\end{table}

\end{document}